\documentclass[a4paper,11pt]{scrartcl}

\makeatletter
\DeclareOldFontCommand{\rm}{\normalfont\rmfamily}{\mathrm}
\DeclareOldFontCommand{\sf}{\normalfont\sffamily}{\mathsf}
\DeclareOldFontCommand{\tt}{\normalfont\ttfamily}{\mathtt}
\DeclareOldFontCommand{\bf}{\normalfont\bfseries}{\mathbf}
\DeclareOldFontCommand{\it}{\normalfont\itshape}{\mathit}
\DeclareOldFontCommand{\sl}{\normalfont\slshape}{\@nomath\sl}
\DeclareOldFontCommand{\sc}{\normalfont\scshape}{\@nomath\sc}
\makeatother

\usepackage[utf8]{inputenc}
\usepackage[T1]{fontenc}

\usepackage[utf8]{inputenc}

\usepackage{multicol}
\addtokomafont{disposition}{\sffamily \mdseries}
\addtokomafont{author}{\large}
\addtokomafont{date}{\sffamily}

\addtokomafont{labelinglabel}{\bfseries}

\usepackage[dvipsnames,svgnames,x11names]{xcolor}

\usepackage{authblk}
\usepackage{array}
\usepackage{graphicx}

\usepackage{booktabs}

\usepackage[english]{babel}

\usepackage[tbtags]{amsmath}
\usepackage{amsfonts}
\usepackage{amssymb}
\usepackage{subfig}

\usepackage{slashed}
\usepackage{graphicx}

\usepackage{dsfont}
\usepackage[scr=dutchcal]{mathalfa}

\usepackage[margin=10pt,textfont={small,it},labelfont=bf]{caption}

\usepackage[numbers,sort&compress]{natbib}

\usepackage{hyperref}

\hypersetup{
        unicode = true,
        pdftitle = Note , 
        colorlinks = true,
        linkcolor = Black!30!DodgerBlue4,
        citecolor = Black!30!DodgerBlue4,
        filecolor = Black!30!DodgerBlue4,
        urlcolor = Black!30!DodgerBlue4,
}

\usepackage[margin=2cm,bottom=3cm,paper=a4paper]{geometry}
\setcapindent{0pt}

\usepackage{booktabs}
\renewcommand{\toprule}{\specialrule{1.5pt}{0em}{1pt} \midrule}
\renewcommand{\bottomrule}{\midrule \specialrule{1.5pt}{1pt}{0em}}
\newcommand{\headrule}{\specialrule{1.5pt}{0em}{0em}}

\graphicspath{{./img/}}

\newcommand{\id}{\mathds{1}}
\newcommand{\unity}{\id}

\newcommand{\ii}{\mathrm{i}}

\renewcommand{\epsilon}{\varepsilon}

\newcommand{\refeq}[1]{eq.~\eqref{#1}}

\newcommand{\reftab}[1]{tab.~\ref{#1}}

\newcommand{\ie}{\textit{i.e.}}

\newcommand{\al}{\alpha}

\renewcommand{\bar}{\overline}

\title{
Dark Matter interactions in an $S_4 \times Z_5$ flavor symmetry framework
}
\author[a,b]{\sffamily G. Arcadi\thanks{giorgio.arcadi@uniroma3.it}}
\author[a]{\sffamily M. B. Krauss}
\author[a]{\sffamily D. Meloni\thanks{davide.meloni@uniroma3.it}}
\affil[a]{\rmfamily Dipartimento di Matematica e Fisica, Università di Roma 
Tre, 
Via della Vasca Navale 84, 00146 Rome, Italy}
\affil[b]{\rmfamily INFN Sezione di Roma Tre}

\begin{document}
% \twocolumn[
% \begin{@twocolumnfalse}

\maketitle

\noindent\hrulefill
\begin{abstract}
\noindent
The interactions of dark matter (DM) with the visible sector are often phenomenologically described in the framework of simplified models where the couplings of quarks to the new particles are generally assumed to be universal or have a simple structure motivated by observational benchmarks. They should, however, a priori be treated as free 
parameters. In this work we discuss one particular realization of the structure of DM couplings based on an $S_4 \times Z_5$ flavor symmetry, which has been shown to account reasonably well for fermion masses and mixing, and 
compare their effect on observational signals to universal as well as Yukawa-like couplings, which are motivated 
by minimal flavor violation. We will also comment on how these structures could 
be constrained in UV complete theories of DM and how DM observables, such as, e.g., relic density and direct detection, can 
potentially be used as a smoking gun for the underlying flavor symmetries.
\end{abstract}
\noindent\hrulefill

% \end{@twocolumnfalse}]

\clearpage

% \begin{multicols}{2}
\setcounter{tocdepth}{1}
\tableofcontents

\section{Introduction}

Motivated by a substantial amount of evidence from cosmological observation, 
there is a continuing effort by experimental collaborations to obtain evidence 
for the existence of particle dark matter~(DM). This effort is supported by a 
large number of theoretical analyses investigating the potential 
characteristics of DM in preparation for a future discovery \footnote{The XENON1T collaboration has recently reported an excess in low energy electron recoil events \cite{Aprile:2020tmw}. The latter cannot, however, be unambiguously interpreted as a DM signal. We will therefore not consider it in this work.}.  There are several options for the choice of framework in which DM can be studied. One can, for example, rely on specific Particle Physics scenarios, like Supersymmetry, and directly confront their predictions with experiments. An orthogonal approach, whose popularity is increased in the recent years, consists of considering simplified models, agnostic to the details of underlying BSM theories, containing only the relevant degrees of freedom, namely the DM and a mediator of its interactions with (typically) SM fermions, and interactions accounting for the DM relic density and possible observable signals at experiments. This kind of approach has been extensively applied to WIMP models \cite{Jacques:2015zha,Xiang:2015lfa,Backovic:2015soa,Bell:2015rdw,Brennan:2016xjh,Boveia:2016mrp,Englert:2016joy,Goncalves:2016iyg,DeSimone:2016fbz,Liew:2016oon,Kraml:2017atm,Bauer:2017ota,Albert:2017onk,Arcadi:2017kky}. As pointed out, e.g. in \cite{Kahlhoefer:2015bea,Bell:2016uhg,Englert:2016joy,Bell:2016ekl,Goncalves:2016iyg}, interpreting the experimental outcome in terms of simplified models requires care in view of their theoretical limitations, as for example the lack of gauge invariance. For this reason the theoretical community is currently working on more solid refinements of simplified models, see e.g. \cite{Ipek:2014gua,Baek:2017vzd,Bauer:2017ota,Bell:2015rdw,Goncalves:2016iyg,Bell:2017rgi,Abe:2018bpo,Arcadi:2017kky,Arcadi:2019lka,Arcadi:2020gge,Arcadi:2020jqf} for discussions. By refinements, we mean models in which the coupling of a SM singlet DM candidate is realized in a renormalizable and possibly gauge invariant way but still maintaining a low number of free model parameters.   

Along this line of reasoning this manuscript aims to investigate the flavor structure of simplified models. For the latter, one typically relies on the simplifying assumption of flavor-diagonal couplings of the mediator. While this assumption is reasonable and elegant, UV complete scenarios might motivate different choices for the assignations of the couplings of the mediator with SM fermions, see e.g.~\cite{Dienes:2013xya,chala:2015ama,DEramo:2016gos}.
Non trivial flavor structures in the fermion mass matrices can be transferred  to the couplings of the SM quarks with the mediator fields, and can be obtained by means of additional symmetries which add to the gauge group  of the SM. In recent years, they have been quite popular thanks to their ability to explain with good accuracy  the observed pattern of neutrino masses and mixing, generally at the prize of introducing heavy scalar degrees of freedom to mediate the symmetry breaking. Among the various realizations, discrete non-abelian groups $G_f$ \cite{Altarelli:2010gt,Ishimori:2010au}  have gained a lot of interest since they have non-trivial multidimensional representations useful to inter-relate different families; in particular, permutation groups as $A_4, S_4...$ were studied in detail as neutrino mixing matrices of tri-bimaximal-mixing \cite{Harrison:2002er} and bimaximal-mixing \cite{Barger:1998ta} forms  naturally emerged from particle assignment to the irreducible tridimensional representations. Quite often such discrete symmetries have been supplemented by $U(1)$ groups a la Froggatt-Nielsen \cite{Froggatt:1978nt} or their discrete counterparts, $Z_N$ groups, ($N$ being the order of the group formed from the  p-th roots of unity) in order to eliminate unwanted operators preventing the correct description of fermion mass hierarchies.
Although not exhaustively discussed in the literature, the compound $G_f \times Z_N$ has been also applied to quarks (see for instance \cite{Altarelli:2005yx,Blum:2007nt,Bazzocchi:2009pv,Ishimori:2014nxa,Feruglio:2007uu,Ma:2002yp,Ding:2008rj}), showing that a leading order (LO) diagonal $V_{CKM}$ can be easily reproduced and higher-order corrections are needed to accommodate the off-diagonal entries.  In the present work we adopt a flavor model based on an underlying $S_4 \times Z_5$ flavor symmetry \cite{Meloni:2009cz} which goes along the lines proposed in \cite{Altarelli:2005yx}, making explicit computation of all next-to-leading order (NLO) terms necessary to describe quark masses and mixing.

Our paper is organized as follows: In section \ref{sec:theoretical_framework} we introduce two low energy models, featuring scalar and fermionic DM coupled with a scalar mediator in turn coupled with SM quarks. The flavor structure of the latter coupling will be inspired by the $S_4 \times Z_5$ symmetry. In section \ref{sec:pheno} the main phenomenological constraints applied to our study will be illustrated and then transferred to the simplified model in section \ref{sec:results1}.
In section \ref{sec:results2}, before the conclusions, we will instead consider the phenomenology of a concrete realization of the model, illustrated in detail in appendix \ref{sec:S4_model}, in which the results are also dependent on a New Physics scale $\Lambda$ associated to the flavor symmetry. Besides the conventional freeze-out, in this last scenario we will consider also the case of DM production through the so-called freeze-in mechanism. The latter allows to achieve the correct relic density for very high values of $\Lambda$, namely $10^{11 \div 12}\,\mbox{GeV}$. In appendix \ref{sec:app_S4} we finally give some general details about the $S_4$ group theory.

\section{Low energy simplified model}
\label{sec:theoretical_framework}
For what concerns Dark Matter and collider phenomenology, the framework under study can be represented as a portal scenario in which an electrically neutral scalar field $\phi$ couples with pairs of SM quarks and pairs of DM particles. For the latter, we will focus, for definiteness, on the cases of scalar $S$ and (dirac\footnote{In the case of a scalar mediator there is no substantial difference between Dirac and Majorana DM. There is therefore no loss of generality in our assumption.}) fermionic $\chi$ DM.

The corresponding lagrangians are given by:
\begin{eqnarray}
\label{eq:lagrangian_scalar}
\mathcal{L}_{S\phi q} &=& \partial_\mu S^\dagger\partial^\mu S - m_S^2S^\dagger 
S - \frac{\lambda_S}{2}(S^\dagger S)^2 \nonumber\\
&+&\frac{1}{2}\partial_\mu\phi\partial^\mu\phi - \frac{1}{2}m_\phi^2\phi^2 
-\frac{m_\phi\mu_1}{3}\phi^3-\frac{\mu_2}{4}\phi^4 \nonumber\\
&+& \ii\bar{q}\slashed{D} q - m_q \bar q q \nonumber\\
&-&g_1m_SS^\dagger S\phi -\frac{g_2}{2}S^\dagger S\phi^2-h_1\bar q 
q\phi-ih_2\bar{q}\gamma^5q\phi\,, \label{eq:Sphi}
\end{eqnarray}
and
\begin{eqnarray}
\label{eq:lagrangian_fermion}
\mathcal{L}_{\chi\phi q} &=& \ii\bar{\chi}\slashed{D}\chi - 
m_{\chi}\bar{\chi}\chi \nonumber\\
&+&\frac{1}{2}\partial_\mu\phi\partial^\mu\phi - \frac{1}{2}m_\phi^2\phi^2 
-\frac{m_\phi\mu_1}{3}\phi^3-\frac{\mu_2}{4}\phi^4 \nonumber\\
&+& \ii\bar{q}\slashed{D} q - m_q \bar q q \nonumber\\	
&-&\lambda_1\phi\bar{\chi}\chi 
-i\lambda_2\phi\bar{\chi}\gamma^{5}\chi-h_1\phi\bar q 
q-ih_2\phi\bar{q}\gamma^5q\,,\label{eq:phichi}
\end{eqnarray}
for scalar and fermionic DM, respectively. In both lagrangians $h_1=h_2$ represent $3 \times 3$ matrices in the quark flavor space. Since we are considering the mediator $\phi$ coupled with all the six SM quarks, there are two copies of $h_1$ and $h_2$ associated, respectively, with u-type and d-type quarks. The structure of the matrix elements is specified  by the $S_4 \times Z_5$ symmetry \cite{Meloni:2009cz} (discussed in details in Appendix \ref{sec:S4_model}) which, at the leading order, are:
\begin{equation}
\label{hleadingorder}
h_1=\mbox{diag}(h_1^d,h_1^s,h_1^b)=\mbox{diag}(\epsilon^3,\epsilon^2,\epsilon) \,,
\end{equation}
with $\epsilon=0.05$ (and ${\cal O}(1)$ coefficients  simply fixed to unity). The same definition applies also for the matrix acting on the u-type quarks. As a further assumption, we will take, throughout our study, $\mu_1=0$, for both fermionic and scalar DM.

Two notable differences appear with respect to the simplified models usually considered in the literature. First of all the couplings $h_1,h_2$ should be intended as $3 \times 3$ matrices whose structure is determined by a specific UV model (see next subsection for more detail on this realization) and should comply with specific observational constraints from processes related to flavor physics. A further important feature is represented by the fact that the lagrangians are explicitly CP violating. This is required to properly account for the flavor structure of the SM, and has also relevant phenomenological implications for DM.

The coupling of the mediator with SM quarks originates, at the loop level, an effective coupling with gluons and photons, relevant for collider phenomenology. The latter can be described through the following lagrangian:
\begin{equation}
\mathscr{L}_{gg\phi} = \tilde h_1^g \phi G_{\mu\nu}G^{\mu\nu} + \ii \tilde h_2^g \phi G_{\mu\nu}\tilde{G}^{\mu\nu}+\tilde h_1^\gamma \phi \, 
F_{\mu\nu}F^{\mu\nu}+\ii \tilde h_2^\gamma \phi \, F_{\mu\nu}\tilde{F}^{\mu\nu} \,,
\label{eq:eff_gluon_couplings}
\end{equation}
where~\cite{Baum:2017gbj}:
\begin{align}
& \tilde h_1^g = \frac{\sqrt{2}\alpha_S}{6\pi v} \frac{3}{4} \sum_q 
h_1^q A_q^S(\tau_q) \,\,\,\,\,\, \tilde{h}_1^\gamma = \frac{\sqrt{2}\alpha}{8\pi v} \frac{3}{4} \sum_q Q_q^2
h_1^q A_q^S(\tau_q),\nonumber\\
& \tilde h_2^g = \frac{\sqrt{2}\alpha_S}{4\pi v} \sum_q 
h_2^q A_q^P(\tau_q) \,\,\,\,\,\,\,\,\, \tilde h_2^\gamma = \frac{3\sqrt{2}\alpha}{16\pi v} \sum_q Q_q^2 
h_2^q A_q^P(\tau_q).
\end{align}
Here $A_q^S$ and $A_q^P$ are loop functions given by:
\begin{align}
 A_q^S(\tau) &= 2 [ \tau + (\tau-1) f(\tau)] / \tau^2 \\
 A_q^P(\tau) &= f(\tau)/\tau \\
 \intertext{where}
 f(\tau) &= \left\{ \begin{matrix} 
    \arcsin^2 \sqrt{\tau} & \tau \leq 1 \\
    -\frac{1}{4} \left[ \log \frac{1+\sqrt{1-1/\tau}}{1-\sqrt{1-1/\tau}} - 
\ii\pi \right]^2 & \tau > 1
 \end{matrix} \right.\,,
\end{align}
and
\begin{equation}
 \tau = \frac{\hat s}{4m_q^2}\,,
\end{equation}
with $\hat s$ being the gluon--gluon center of mass energy.

\section{Phenomenological constraints}
\label{sec:pheno}

\subsection{Relic Density}
\label{sec:thermal_production}

Throughout this paper we will mostly consider the standard thermal freeze-out paradigm for the determination of the DM relic density. According to it, the latter is determined by the thermally averaged DM pair annihilation cross-section $\langle \sigma v \rangle$. In the scenario under consideration DM annihilates into SM quark pairs, through s-channel exchange of the mediator $\phi$, and into $\phi \phi$, through t-channel exchange of the DM (in the case of scalar DM a contact interaction vertex is present as well), if kinematically allowed \footnote{Notice that the relic density phenomenology can be altered in more realistic setups in which additional BSM states are present. See e.g. \cite{Arcadi:2016kmk,Arcadi:2016qoz}.}. For completeness, we have also included in our analysis the annihilation channels into $gg$ and $\gamma \gamma$ originated by the Lagrangian in  \ref{eq:eff_gluon_couplings}. Useful analytical expressions can be found e.g. in \cite{Berlin:2014tja,DEramo:2016aee,Catena:2017xqq,Arcadi:2017kky,Arcadi:2019lka}. Our results will be, however, obtained through precise numerical computations as performed by the package micrOMEGAs \cite{Belanger:2006is,Belanger:2007zz}.

%\section{Impact on direct detection signals}
\subsection{Direct Detection}
\label{sec:direct_detection}

The flux of WIMP DM through a detector volume on Earth can lead to scattering events in a suitable target. This in turn can be detected as recoil energy. This kind of phenomena is at the base of the so-called DM Direct Detection (DD).

For the models under consideration, the DM scattering processes are described, at the microscopic level, by interactions of DM pairs with quark (and gluon) pairs through t-channel exchange of the mediator $\phi$. These microscopic interactions lead mostly to Spin-Independent interactions between the DM and the nucleons whose corresponding cross-section is given by:
\begin{equation}
    \sigma_{\rm DM \, N}^{\rm SI}=\frac{\mu_{\rm DM, N}^2}{\pi}|c_{\rm DM, N}|^2\,,
\end{equation}
where $\mu_{\rm DM,N}$ is the DM-nucleon reduced mass while $c_{\rm DM,N}$ is the effective DM-nucleon coupling. In the case of scalar DM the latter is given by:
\begin{equation}
     c_{S,N}=m_N \sum_{q=u,d,s} \frac{f^N_{Tq}}{m_q}\left(\frac{h_1^q g_1}{ m_{\phi}^2}\right)+\sum_{q=c,b,t}\frac{2}{27} m_N f_{TG}^N \frac{g_1 h_1^q}{2 m_q m_\phi^2}\,,
\end{equation}
where $f_{Tq}$ are structure functions describing the contribution of the light quarks to the nucleon mass. 
For these we have adopted the following numerical values \cite{Crivellin:2013ipa,Junnarkar:2013ac,Hoferichter:2015dsa} :
%\textbf{si potrebbe aggiungere una referenza a questi numeri?}:
\begin{align}
& f_{Tu}^p=(20.8 \pm 1.5)\times 10^{-3},\,\,\,\, f_{Tu}^n=(18.9\pm 1.4) \times 10^{-3} \\
& f_{Td}^p=(41.1 \pm 2.8)\times 10^{-3}\,\,\,\,\,\;\,f_{Td}^n=(45.1 \pm 2.7)\times 10^{-3} \\
& f_{Ts}^p=0.043\pm 0.011\qquad \;\;\;\;\;\;\,  f_{Ts}^n=0.043\pm 0.011\,.
\end{align}
The coefficient $f_{TG}^N$ is instead given by $f_{TG}^N=1-\sum_q f_{Tq}^N$. In the case of fermionic DM the effective coefficient is a combination of a tree-level and loop-induced contributions~\cite{Arcadi:2017wqi,Abe:2018emu,Sanderson:2018lmj,Ertas:2019dew}: 
\begin{align}
   & c_{\chi\,N}=\sum_{q=u,d,s}m_N \frac{f_{Tq}^N}{m_q}\left(\frac{h_1^q \lambda_1}{m_{\phi}^2}+C_{1,q}^{\rm box}\right)+\sum_{q=c,b,t}\frac{2}{27} m_N f_{TG}^N \frac{\lambda_1 h_1^q}{m_q m_\phi^2}\nonumber\\
    & +\frac{3}{4}m_N m_\chi \left[q(2)^N+\bar{q}^N (2)\right] \left(C_{5,q}^{\rm box}+m_\chi C_{6,q}^{\rm box} \right)+\frac{2}{27}m_N f_{TG}^N C_{G,S}^{\rm eff}\,.
\end{align}
The quantities $C_{i,q}^{box}$ and $C_{G (\tilde{G}),S (PS)}^{\rm box}$ are loop functions which depend on the model parameters, i.e. $m_{\chi,\phi}$, $\lambda_{1,2}$ and $h_{1,2}$. Since these expressions are rather complicated, we do not report them here explicitly. The interested reader can find them in \cite{Abe:2018emu,Ertas:2019dew}.

\subsection{Indirect Detection}

All the scenarios considered in present work are characterized by DM annihilation cross-sections with sizable s-wave component. Consequently, we can have efficient residual annihilation processes at present times which can be probed through DM Indirect Detection (ID) search strategies. However, while this is a generic feature of the case of scalar DM, for fermionic DM the prospects for ID rely only on the pseudoscalar coupling $\lambda_2$. This is because the scalar coupling $\bar \chi \chi \phi$ is responsible of a velocity dependent (p-wave) contribution to the annihilation cross-section, whose value at present times is hence suppressed with respect to the one at thermal freeze-out. The pseudoscalar coupling $\bar \chi \gamma_5 \chi \phi$ leads, instead, to an s-wave contribution to the DM annihilation cross-section. 
DM annihilations into SM quarks lead to a $\gamma$-ray signal with continous energy spectrum, originating in the quark hadronization process (e.g. decay $b \rightarrow \pi^0 \gamma$). This kind of signal can be probed, for the range of DM masses considered in our work, by the FERMI-LAT experiment~ \cite{Clark:2017fum,Ahnen:2016qkx}. Current constraints exclude cross-sections of the order of the thermally favoured value for DM masses $\lesssim 150\,\mbox{GeV}$. In addition to the just illustrated continuous $\gamma$-ray signal, mono-energetic (lines) $\gamma$-rays can be produced as well by $DM DM \rightarrow \gamma \gamma$ processes, made possible by the loop induced coupling of the mediator $\phi$ with a photon pair. We have adopted the constraints presented in~\cite{Ackermann:2015lka}. As shown in figs.~\ref{fig:scalar2D}-\ref{fig:fermion2D}, the latter are nevertheless subdominant.

\subsection{Collider searches}
\label{sec:collider}

Given the lagrangians~(\ref{eq:lagrangian_scalar},\ref{eq:lagrangian_fermion},\ref{eq:eff_gluon_couplings}), potential collider signals originate from the resonant production of the mediator $\phi$ through gluon fusion or quark-quark fusion. The latter process has, however, negligible impact for the assumed flavor structure and the assignation $\epsilon=0.05$. Subsequent decays into visible states lead mostly to a djiet or diphoton signal while, in the case of sizable decay branching fraction into DM, a monojet plus missing energy signal would be originated. 
The model under consideration cannot, however, be probed through the collider searches mentioned before. As shown, e.g. in \cite{Aaboud:2017phn,Sirunyan:2017hci,Aaboud:2019yqu,Alanne:2020xcb}, current results can probe production of mediators with couplings bigger than the order of the SM Yukawa couplings; the top Yukawa coupling $y_t$ is, in particular, crucial to have a sizable production vertex. In the case of $S_4 \times Z_5$ couplings, with $\epsilon=0.05$, top and bottom quark loops equally contribute to the production vertex. Given that, nevertheless, $\epsilon \ll y_t$, the production cross-section of the $\phi$ state is sensitively more suppressed with respect to the Yukawa simplified model.   

%On the contrary, in our model specified in eq.(\ref{hleadingorder}), the coupling of the mediator with the top is of order $\epsilon=0.05$, much lower than the Yukawa coupling of the top.

\section{Results for the simplified model}
\label{sec:results1}

\begin{figure}
    \centering
    \subfloat{\includegraphics[width=0.35\linewidth]{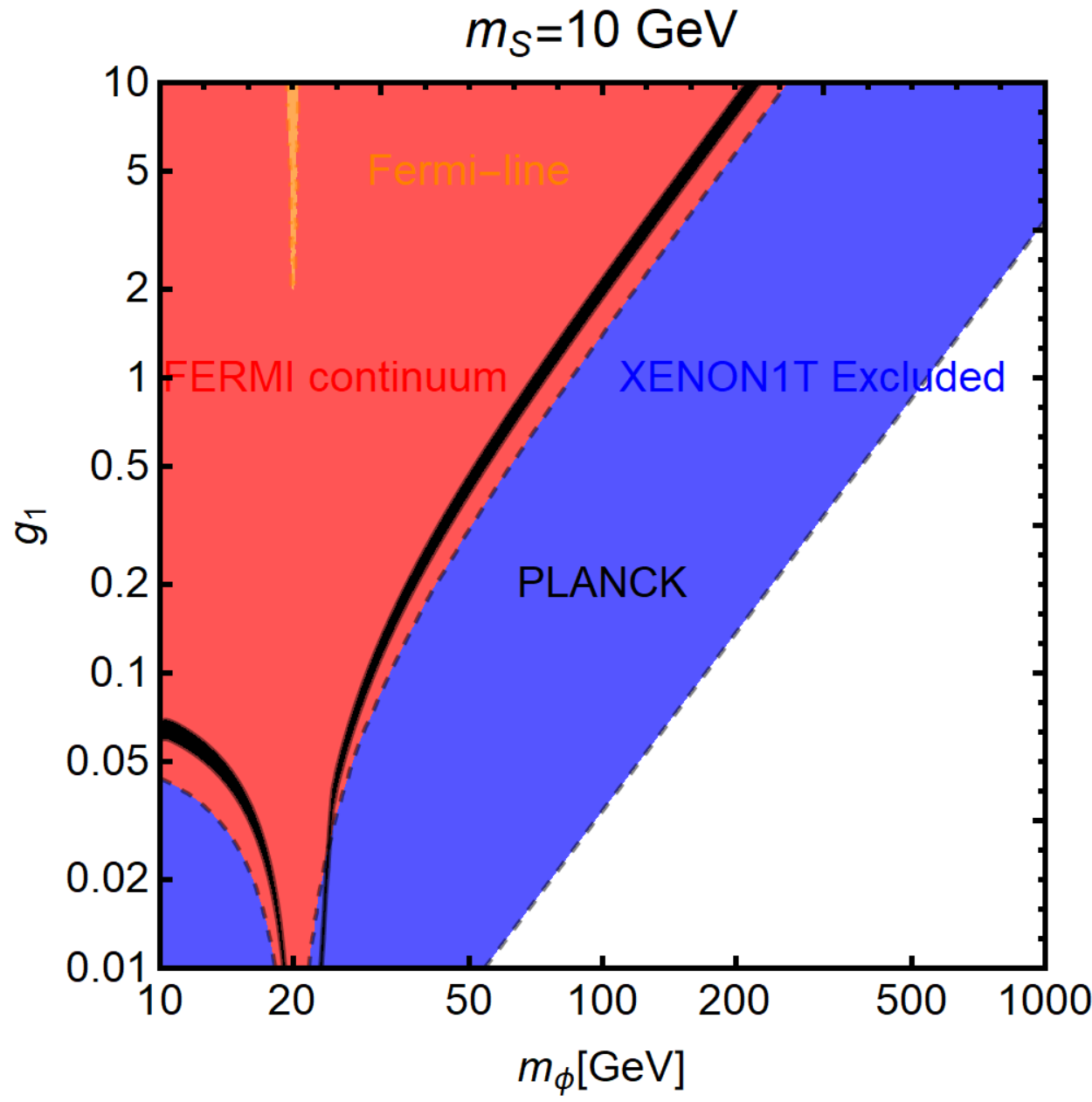}}
    \subfloat{\includegraphics[width=0.35\linewidth]{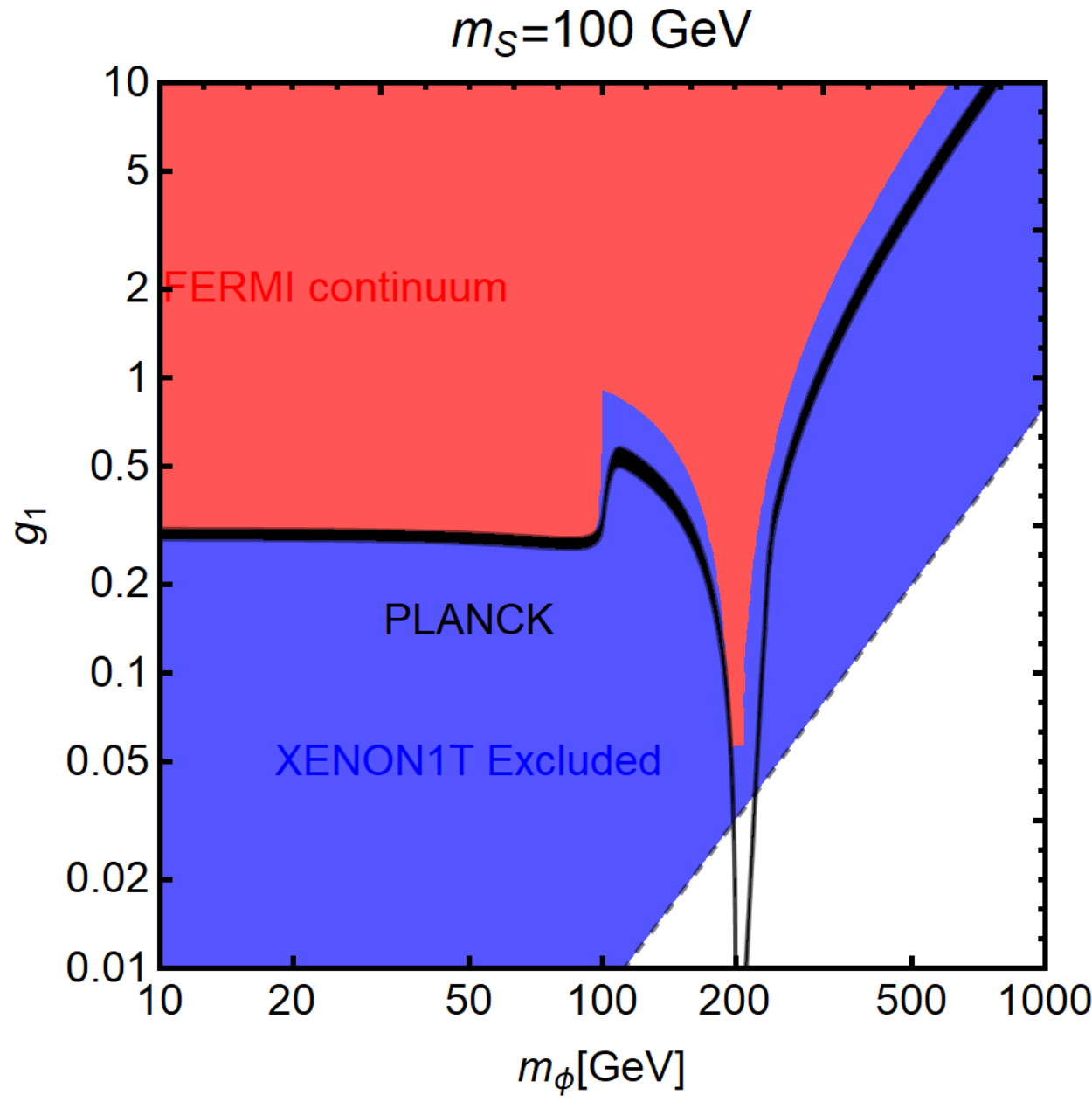}}
    \subfloat{\includegraphics[width=0.35\linewidth]{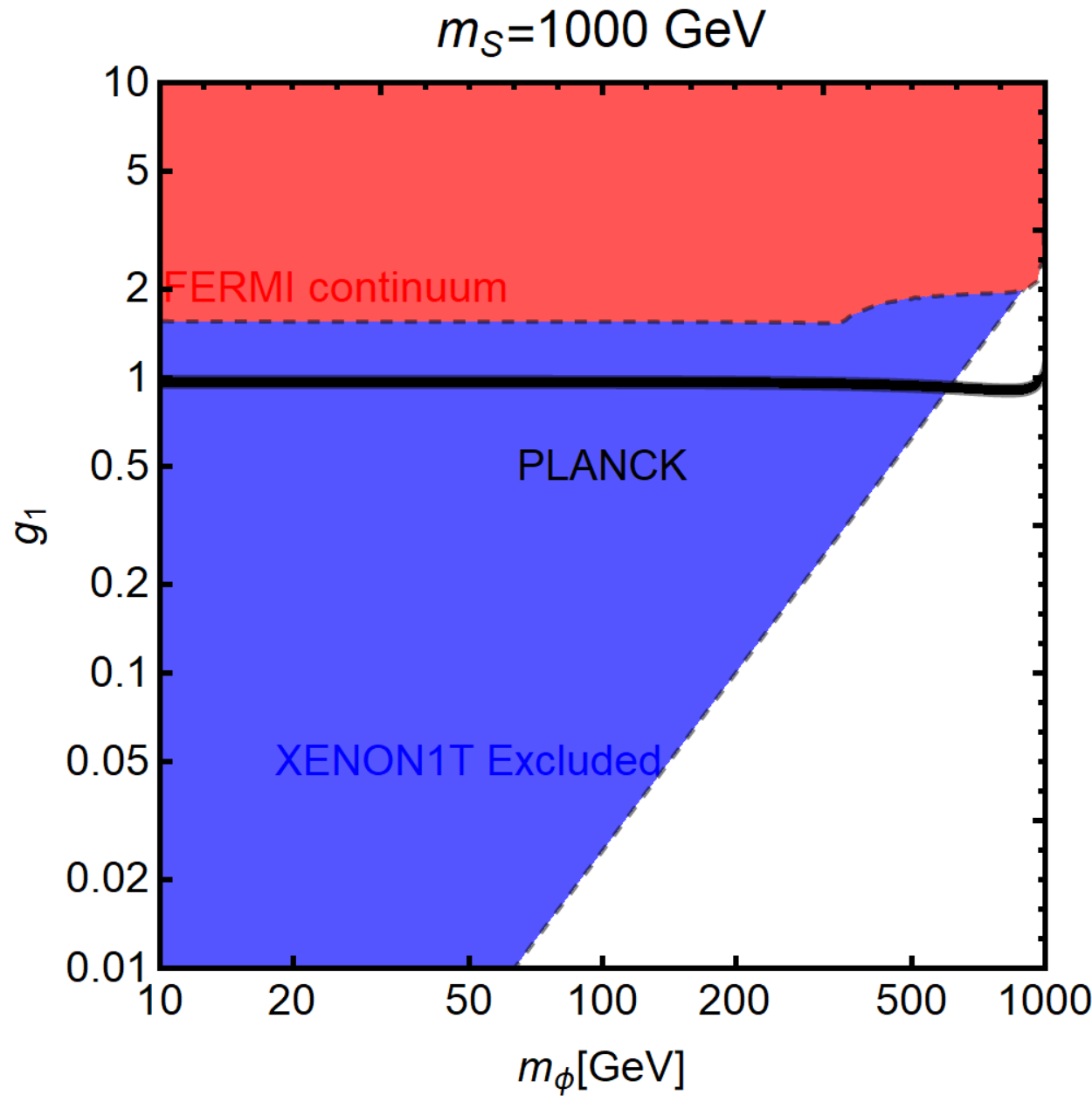}}
    \caption{Combined constraints in the $(m_\phi,g_1)$ bidimensional plane for the three assignations $m_S=10,100,1000\,\mbox{GeV}$ of the DM mass. In all cases we have set $g_2=0$. The black isocontours represent the correct relic density according the standard thermal paradigm. The blue region represents the current exclusion from the XENON1T experiment. The red (orange) regions are the exclusions from searches of continuum (line) $\gamma$-ray signals from DM annihilations at present times.}
    \label{fig:scalar2D}
\end{figure}

\begin{figure}
    \centering
    \subfloat{\includegraphics[width=0.4\linewidth]{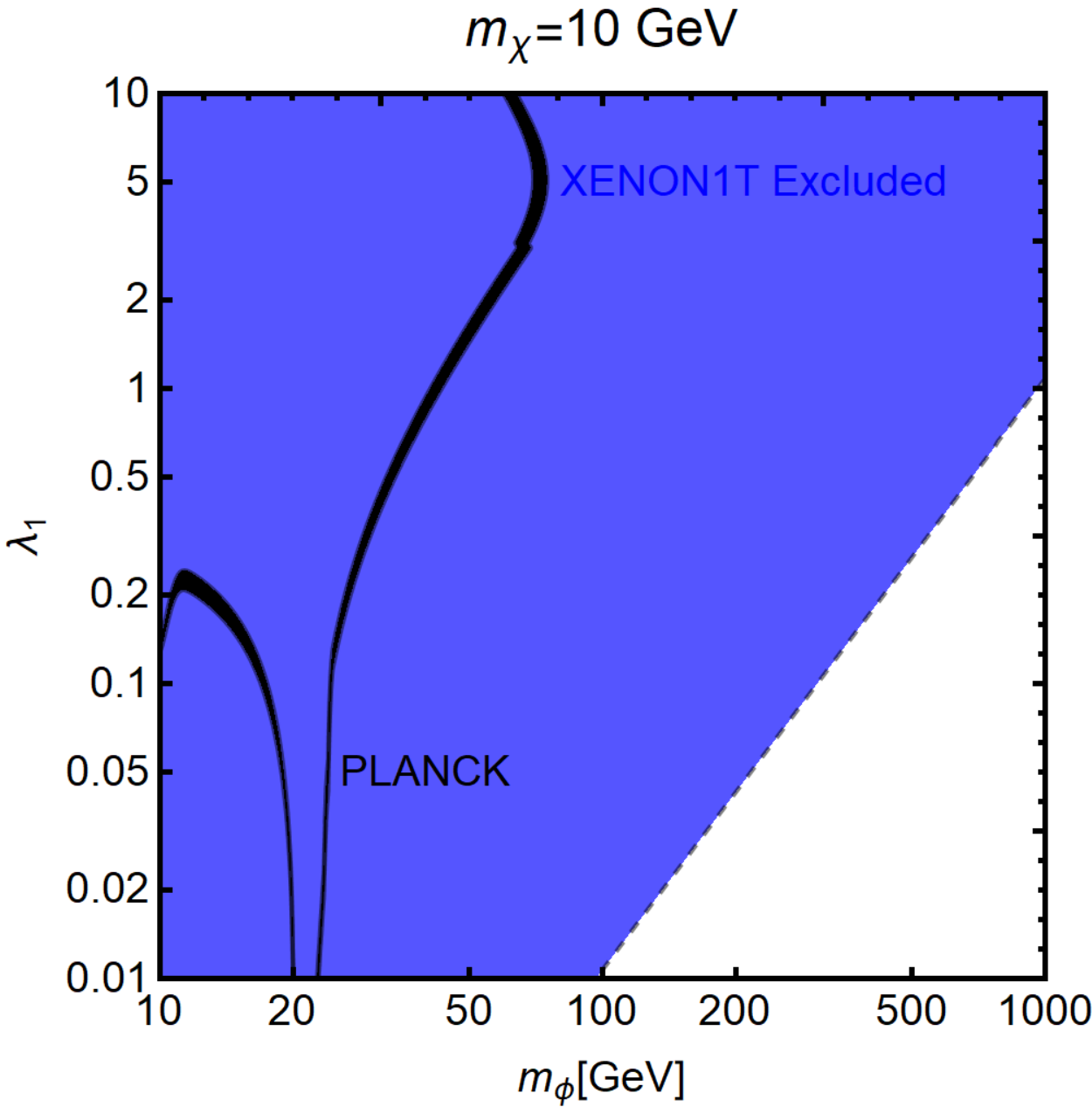}}
    \subfloat{\includegraphics[width=0.4\linewidth]{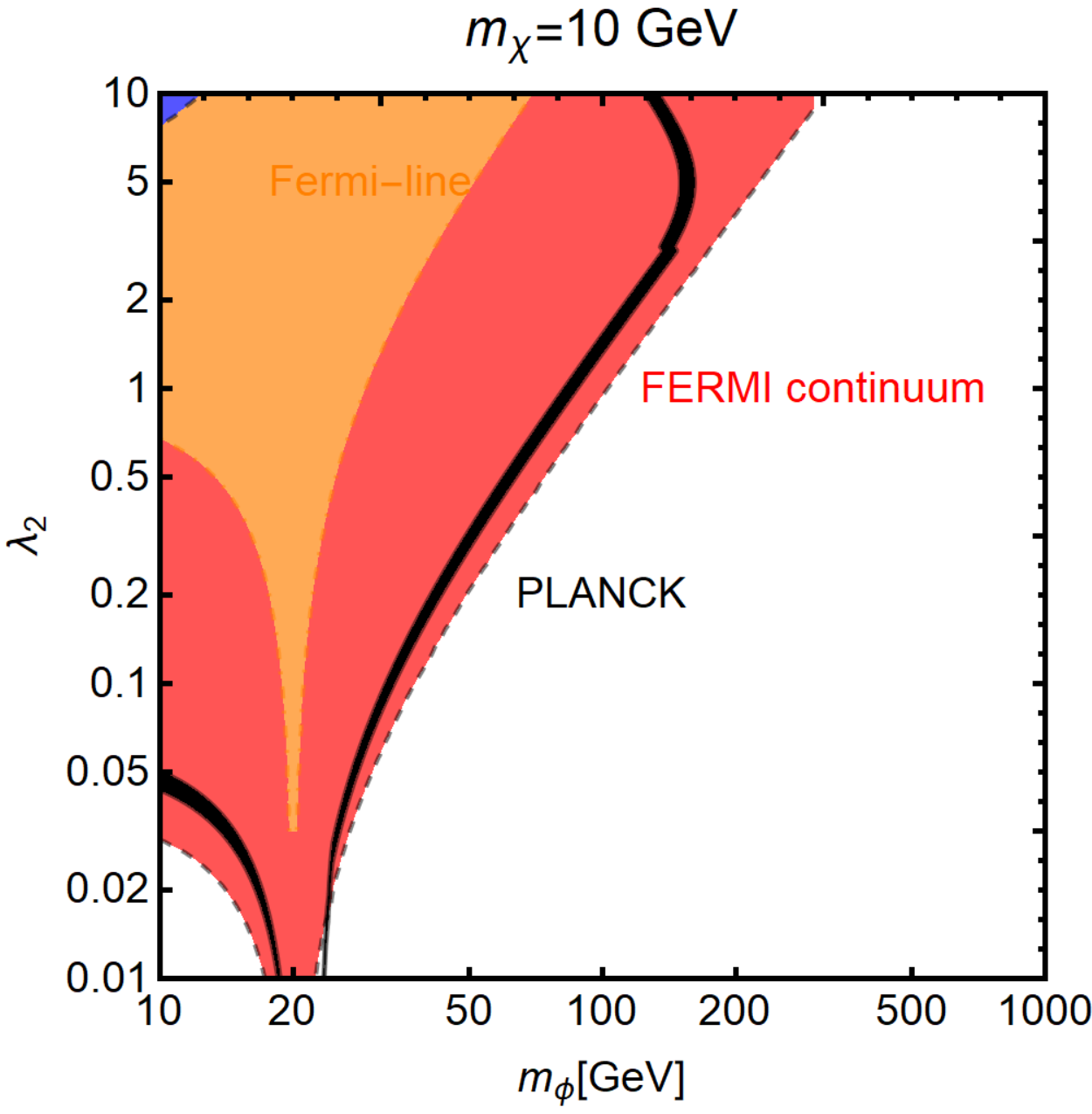}}\\
    \subfloat{\includegraphics[width=0.4\linewidth]{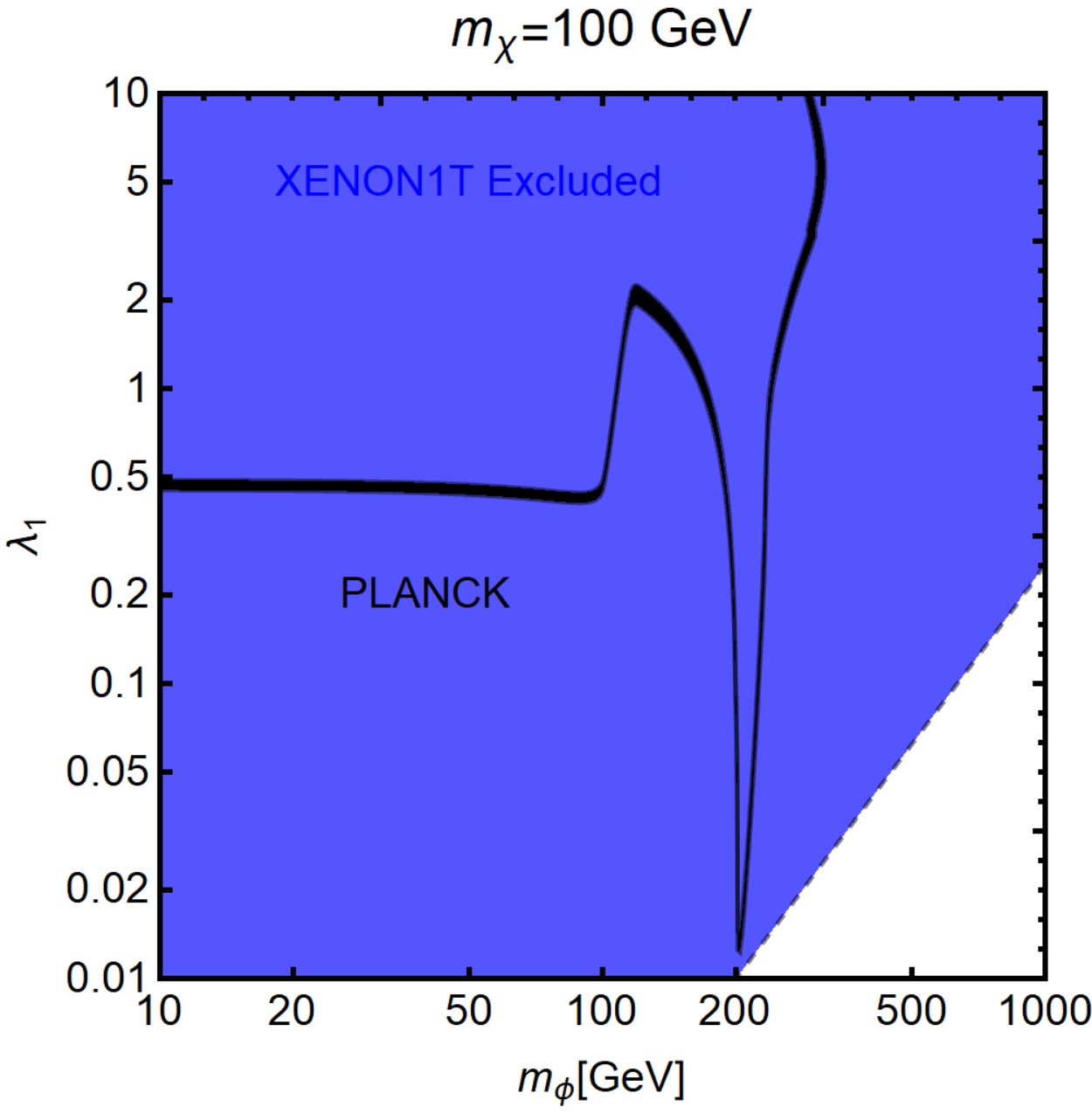}}
    \subfloat{\includegraphics[width=0.4\linewidth]{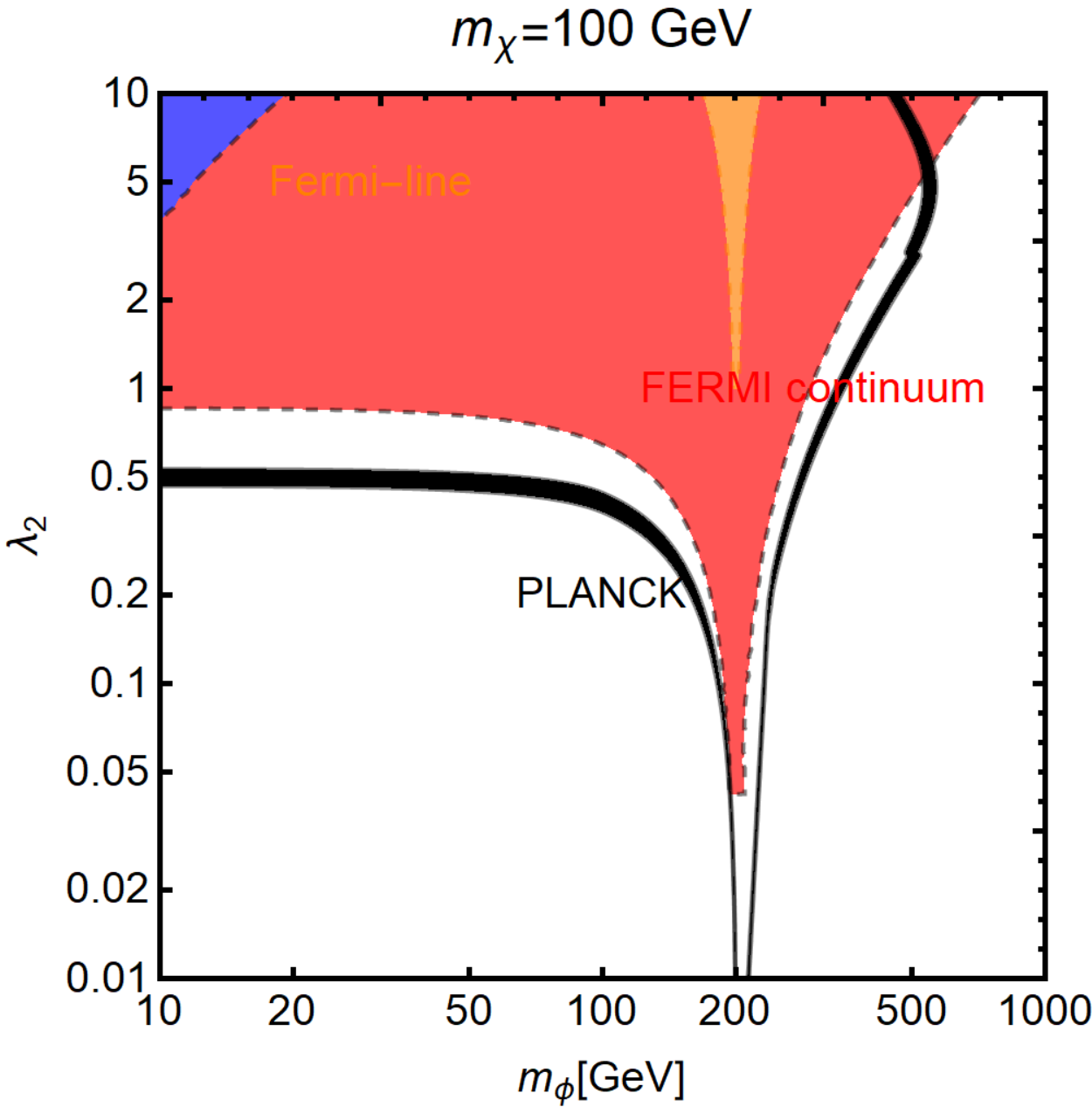}}\\
    \subfloat{\includegraphics[width=0.4\linewidth]{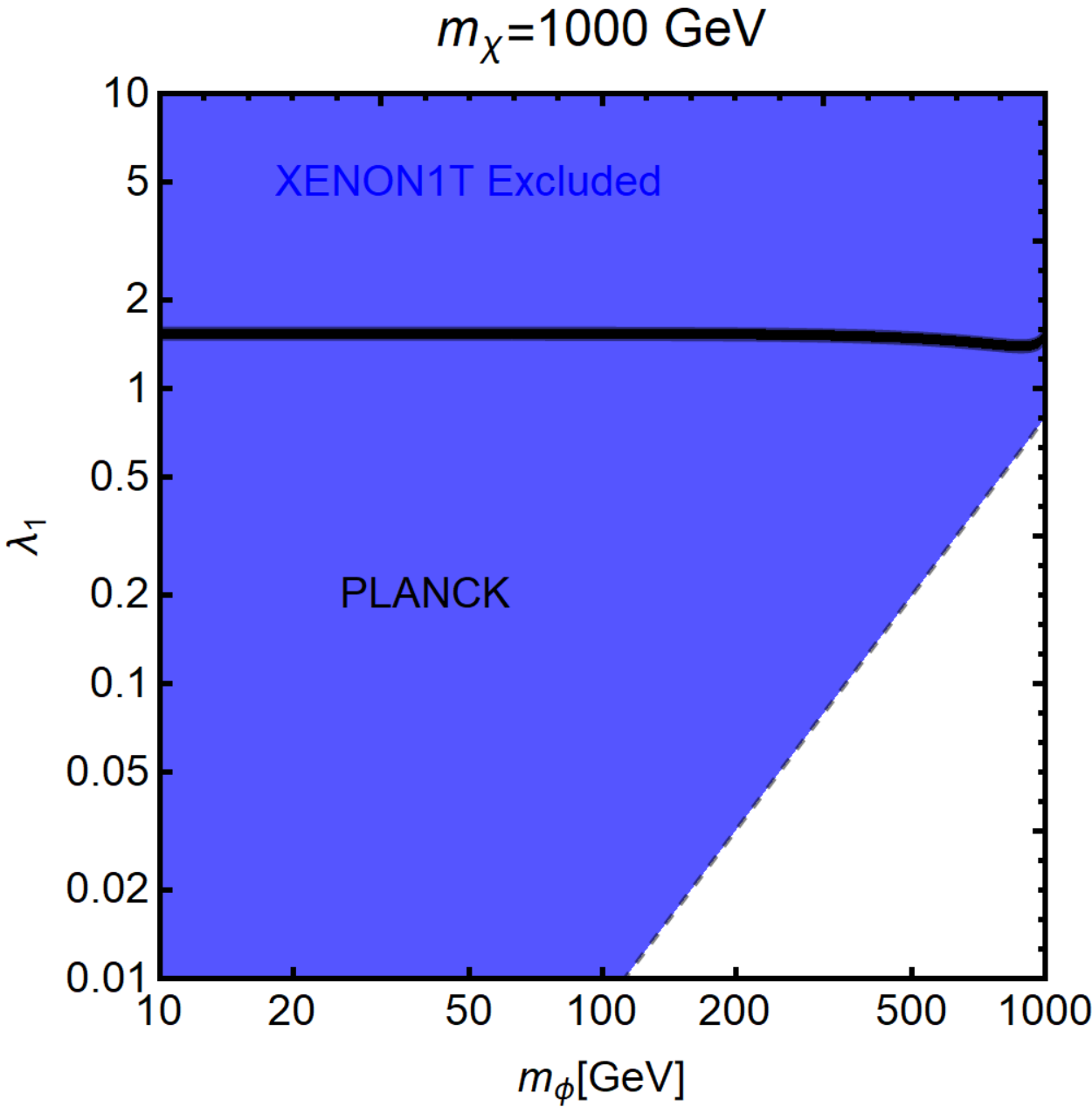}}
    \subfloat{\includegraphics[width=0.4\linewidth]{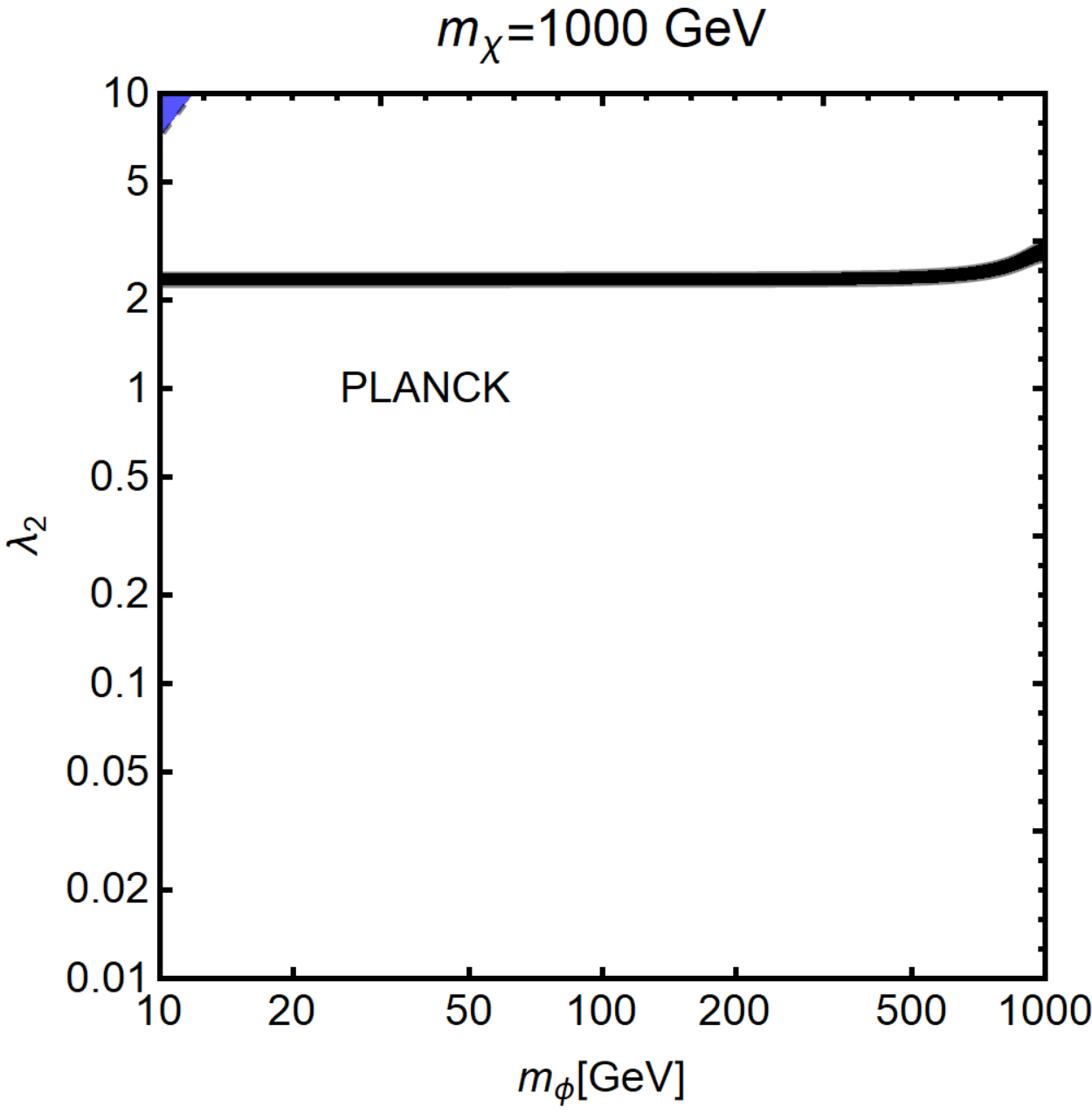}}
    \caption{Same as fig.\ref{fig:scalar2D} but in the $(m_\phi,\lambda_1)$ (left column) and $(m_\phi,\lambda_2)$ (right column). The color code is the same as fig.\ref{fig:scalar2D}.}
    \label{fig:fermion2D}
\end{figure}

We have now all the main ingredients to characterize DM phenomenology within the model specified by the couplings in eq.~(\ref{hleadingorder}). As a first illustration of our results, we have shown in fig.~\ref{fig:scalar2D} and fig.~\ref{fig:fermion2D} the combination of DM constraints in the bidimensional (mass of the mediator, coupling)-planes, for three different assignations of the mediator mass, namely 10, 100 GeV and 1 TeV. In all the plots the parameter spaces corresponding to the correct DM relic density is represented by black isocontours, the regions excluded by DM DD are marked in blue while the regions excluded by indirect searches of DM annihilations in $\gamma$-ray continuum (lines) have been marked in red (orange).

In the case of scalar DM we have considered the $(m_S,g_1)$ bidimensional plane, with the other coupling $g_2$ set to zero. 

This kind of scenario is constrained by both Direct and Indirect detection with the former typically giving the most stringent constraints. As evident, the lightest benchmark with $m_S=10\,\mbox{GeV}$ is completely ruled out by the experimental constraints. The latter exclude also most of the parameter space for $m_S=100,\mbox{GeV}$, ad exception of the pole region, i.e. $m_S \simeq \frac{m_\phi}{2}$. In the case $m_S=1\,\mbox{TeV}$,  the DM relic density is mostly accounted for by the $SS \rightarrow \phi \phi$ annihilation process so that the corresponding isocontour substantially coincides with an horizontal line. Bounds from Direct Detection are evaded for $m_\phi \gtrsim 500\,\mbox{GeV}$.

In the case of fermionic DM, shown in fig. \ref{fig:fermion2D}, we considered the $(m_\phi,\lambda_1)$, with $\lambda_2=0$, and $(m_\phi,\lambda_2)$, with $\lambda_1=0$. 
All the considered DM mass assignations are ruled out in the case only the $\lambda_1$ coupling is on. This is so because, if only $\lambda_1 \ne 0$, the DM annihilation cross-section is velocity suppressed. Consequently, different from the scalar DM case, larger values are required to match the thermally favored value. This causes, in turn, much more stringent constraints from DM DD. More interesting is the case in which only the coupling $\lambda_2$ is different from zero. Indeed it contributes to the DM scattering cross-section only at the loop level and, as evidenced by the right column of fig.\ref{fig:fermion2D}, constrains the parameter space to a negligible extent. The coupling $\lambda_2$ is instead sensitive to constraints from Indirect Detection which, however, exclude only the lightest assignation of the mass of the DM. It can be easily argued that, given the presence of the two couplings $\lambda_{1,2}$, with different properties, there is a broader available parameter space, with respect to the case of scalar DM.

\begin{figure}
    \centering
    \subfloat{\includegraphics[width=0.52\linewidth]{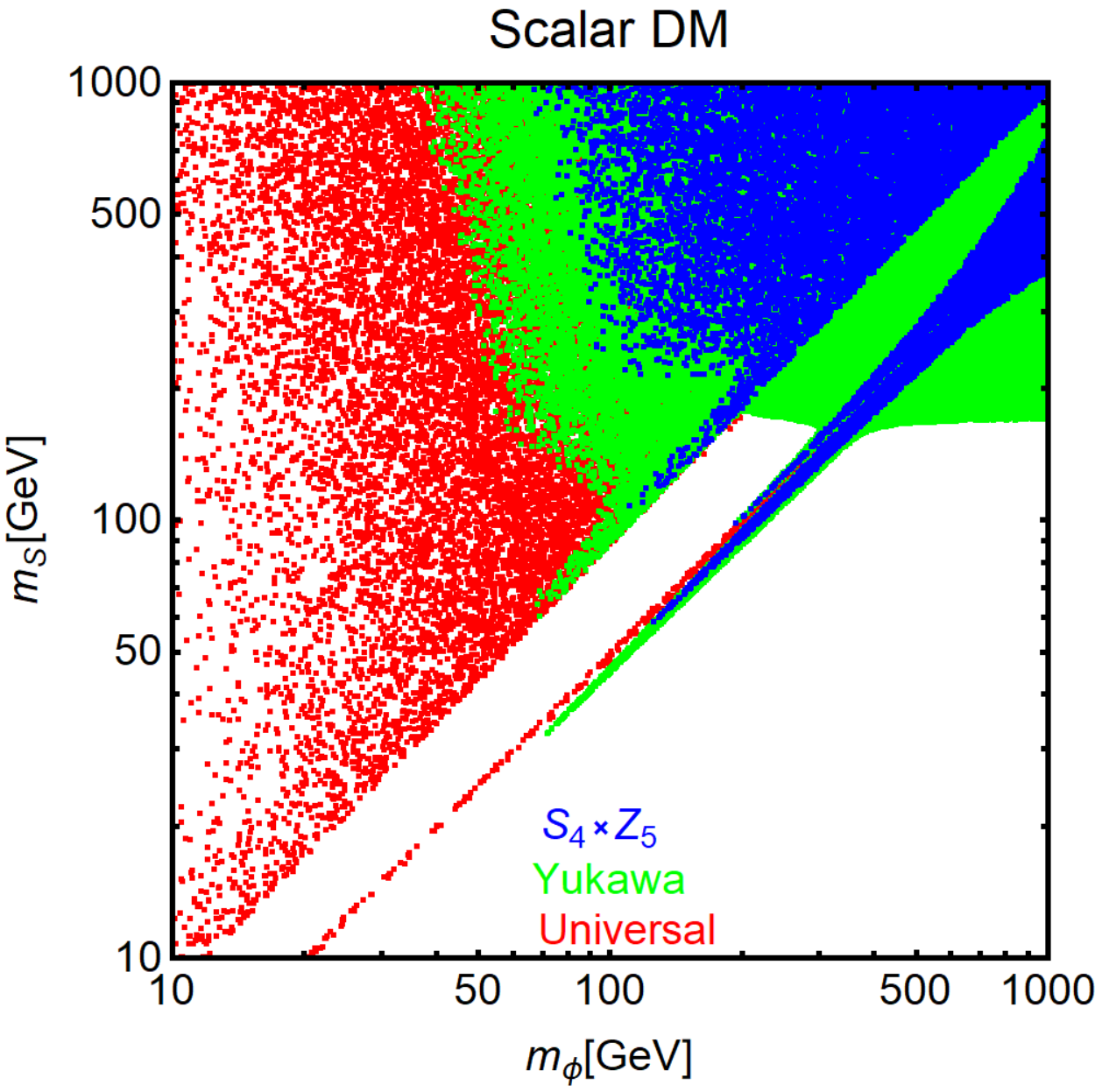}}
    \subfloat{\includegraphics[width=0.5\linewidth]{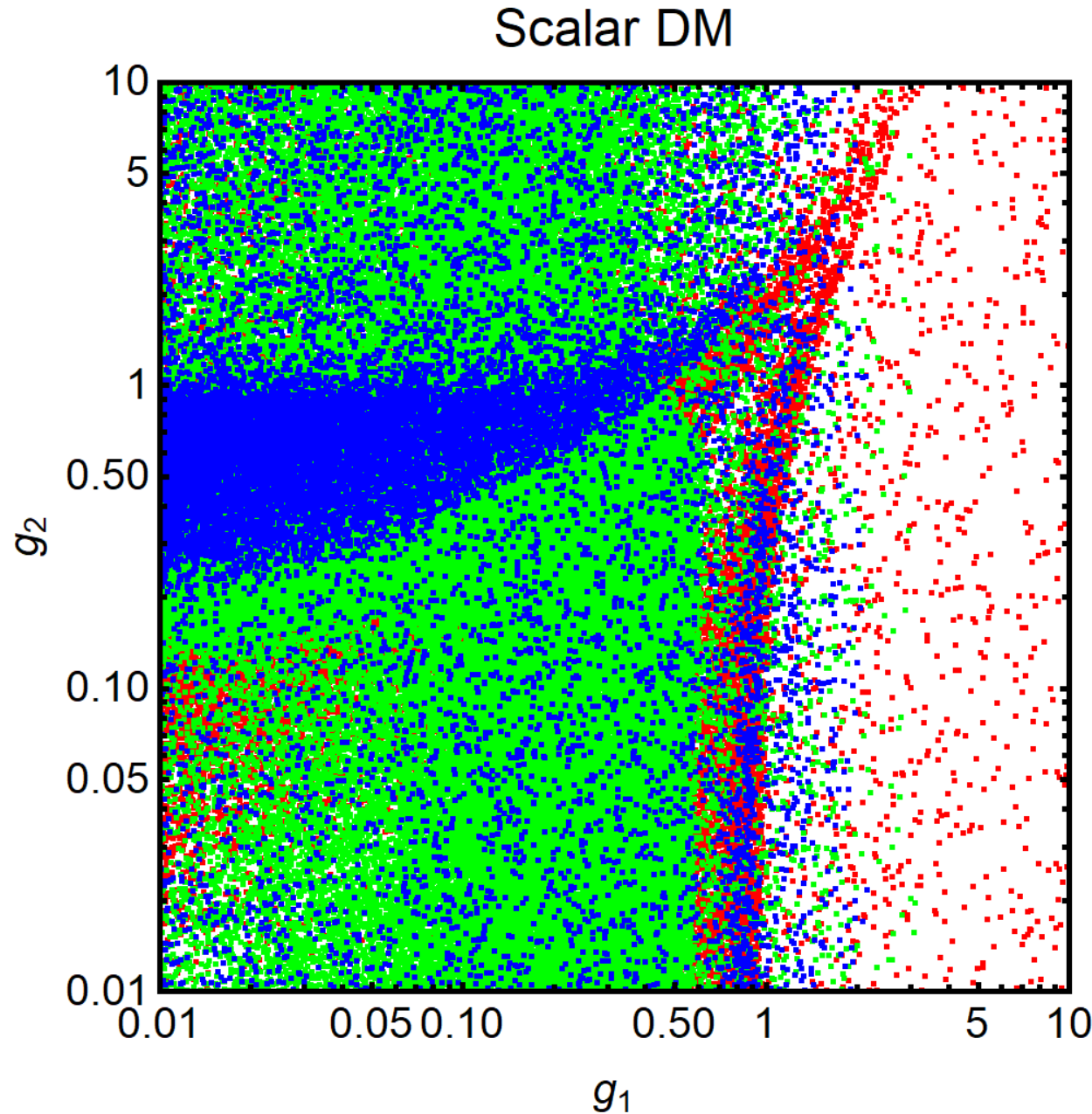}}
    \caption{Viable Model points (marked in blue) in the $(m_\chi,m_\phi)$ (left panel) and $(g_1,g_2)$ (right panel) bidimensional planes for the $S_4 \times Z_5$ inspired simplified model. For comparison the plots show also the results of an analogous study of simplified models with universal (red points) and yukawa-like (green point) couplings between the scalar mediator and the
    quarks.}
    \label{fig:Stot}
\end{figure}

\begin{figure}
    \centering
    \subfloat{\includegraphics[width=0.52\linewidth]{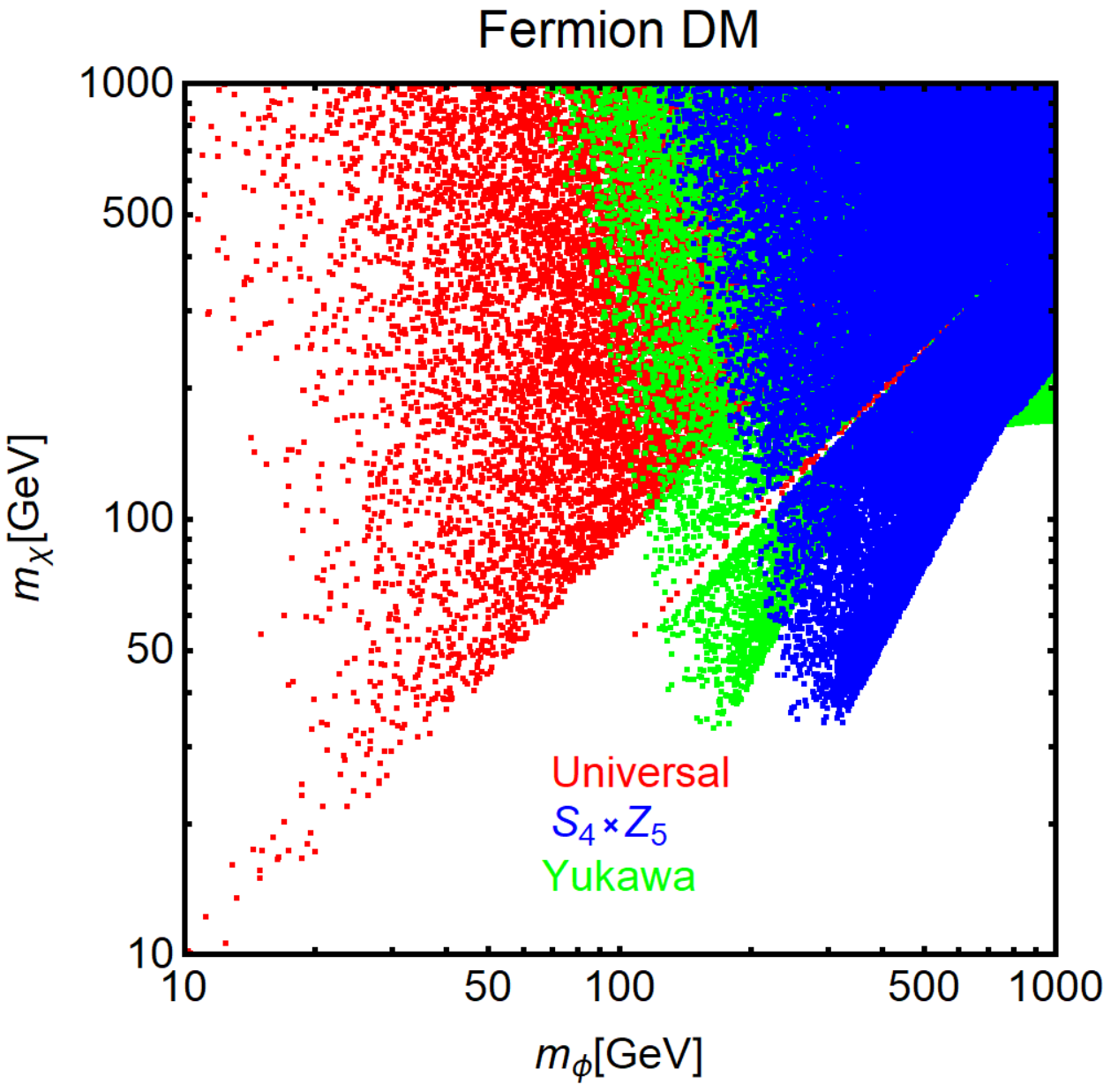}}
    \subfloat{\includegraphics[width=0.5\linewidth]{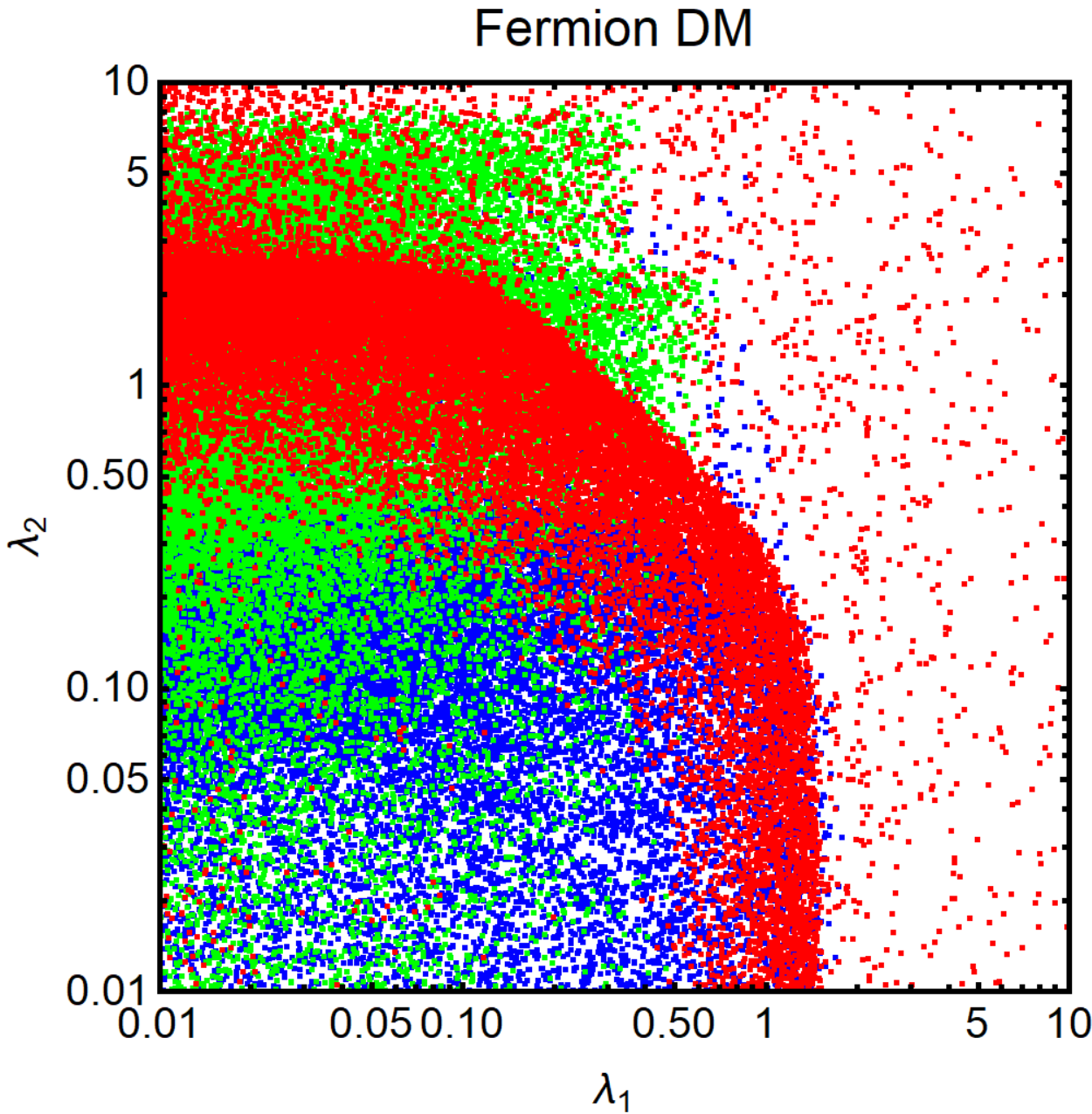}}
    \caption{Same as fig.~\ref{fig:Stot} but for the case of fermionic DM.}
    \label{fig:Ftot}
\end{figure}

Instead of assuming some model parameters fixed to constant values, we made our results more systematic by performing a parameter scan over the following ranges:
\begin{align}
    & m_S \in \left[10,1000 \right]\,\mbox{GeV}\nonumber\\
    & m_\phi \in \left[10,1000\right]\mbox{GeV}\nonumber\\
    & g_1 \in \left[0.01,10\right] \nonumber\\
    & g_2 \in \left[0.01,10\right],
\end{align}

\begin{align}
    & m_\chi \in \left[10,1000 \right]\,\mbox{GeV}\nonumber\\
    & m_\phi \in \left[10,1000\right]\mbox{GeV}\nonumber\\
    & \lambda_1 \in \left[0.01,10\right] \nonumber\\
    & \lambda_2 \in \left[0.01,10\right],
\end{align}
for scalar and fermionic DM, respectively. As already mentioned we have set the tri-scalar coupling $\mu_1$ to zero. The parameter assignations passing all the constraints illustrated in the previous subsections are shown, as blue points, in fig.~\ref{fig:Stot} for scalar DM, and fig.~\ref{fig:Ftot} for fermionic DM. For comparison, the same plots show the results of analogous scans conducted for simplified models with Yukawa-like (green point) and flavor-universal (red points). For Yukawa-like simplified model we intend the case in which~\cite{Tanabashi:2018oca}:
\begin{align}
\label{eq:yukawaSM}
  &  h_1=h_2=\mbox{diag}\,\left(y_u,y_c,y_t\right) \approx (1.03 \times 10^{-5},5.16 \times 10^{-3},0.71);\nonumber\\
  & h_1=h_2=\mbox{diag}\,\left(y_d,y_s,y_b\right) \approx (2.04 \times 10^{-5},4.1 \times 10^{-4},1.70 \times 10^{-2});
\end{align}
for up-type and d-type quarks respectively. $y_{f=u,d,s,c,b,t}$ are the SM yukawa couplings at the EW scale, whose numerical values, have been explicitly reported, for convenience, in eq.\ref{eq:yukawaSM}.

%while $c_\phi$ is common factor which has been varied between 0.01 and 10. 
By quark-universal model we intend, instead, the assignation:
\begin{equation}
    h_1=h_2= c_\phi \mathcal{I}\,,
\end{equation}
with $\mathcal{I}$ being the identity matrix in the flavour space. $c_\phi$ is common factor which has been varied between $10^{-6}$ and 1 (the reason for this choice of the range of the scan will be clarified below). 

Focusing, for the moment, on the $S_4 \times Z_5$ model, we see from fig.~\ref{fig:Stot} that the scalar DM scenario is rather constrained. In agreement with the findings of fig.~\ref{fig:scalar2D}, viable model points are found only for $m_S > m_\phi$ or around the $m_S \simeq m_\phi/2$ ``pole''. Furthermore, only model points with $m_\phi \gtrsim 100\,\mbox{GeV}$ can comply with all the phenomenological constraints. In the case of fermionic DM, on the contrary, we notice viable solutions also for $m_\chi < m_\phi/2$. This is due to the fact that the $\lambda_1$ coupling influences mostly DD while it has negligible impact on ID; on the contrary the $\lambda_2$ coupling impacts mostly ID. It is then possible to evade experimental constraints by a suitable combination of these two couplings.   

Let's now compare the results for the $S_4 \times Z_5$ model with the flavour universal and yukawa cases. In the former case we see that, for both scalar and fermionic DM, there are viable regions of parameters space only for $m_{\chi,S}> m_{\phi}$. This is due to the fact that, in the case of universal couplings, the contributions to the DD cross-section from quarks of the first generations are dramatically enhanced, since they are proportional to $1/m_q$. This does not occur in the other two models since they feature automatically suppressed couplings of the mediator with the first two quark generations. Experimental constraints can be passed only for $c_\phi \ll 0.01$. This implies, in turn, that annihilations into SM fermions are too suppressed to ensure the correct DM relic density, which can be achieved only for $m_{\chi,S} > m_\phi$, when the annihilations into $\phi \phi$ are kinematically accessible \footnote{Notice that the DD cross-section is mostly sensitive to the $h_1$ coupling. The allowed parameter space could be then broadened by taking $h_1 \neq h_2$.}. This kind of scenario is often dubbed {\it secluded regime}.
More interesting is the comparison between the yukawa and the $S_4 \times Z_5$ models. 
In the case of scalar DM we clearly see that the yukawa model have a larger allowed parameter space. There are, in particular, viable solutions for $m_S < m_\phi$ far from the $m_\phi/2$ resonance. This is due to the opening, at high DM masses, of the annihilation channel into $\bar t t$ final state. The same does not occur in the case of the $S_4 \times Z_5$ model since $\epsilon \ll y_t$. Less trivial is, instead, the comparison in the case of fermionic DM. The viable parameter space for the Yukawa simplified model extends at slightly lower mediator masses with respect to the  $S_4 \times Z_5$ case. The former model appears to be also slightly more favourable for $m_\chi > m_t$. This can be explained with the fact that in the $S_4 \times Z_5$ model the mediator $\Phi$ has slightly stronger couplings with the first generation quarks as well as the bottom quarks, which then face stronger experimental constraints. At the same time, the stronger coupling with the bottom, with respect to the yukawa case, opens a viable region or parameter space for $35 \lesssim m_\chi \lesssim 120\,\mbox{GeV}$ and $m_\phi \gtrsim 300\,\mbox{GeV}$.

\section{Towards a more realistic scenario}
\label{sec:results2}

Up to now we have considered the phenomenology of a simplified model in which the couplings between the mediator $\phi$ and the SM quarks are inspired by the structure of a specific flavor group. A more realistic realization can be obtained, in a similar spirit as~\cite{Alanne:2020xcb}, by extending, through two singlet fields, namely the DM candidate and the mediator $\phi$, an $S_4 \times Z_5$ invariant realization of the SM. The details of the model will be left to appendix~\ref{sec:S4_model}. The most relevant impact on DM phenomenology will be due to the dependence of the couplings $h_{1,2}$ to the NP scale $\Lambda$ associated to the breaking of the flavor symmetry:
\begin{equation}
    h_{1,2}\rightarrow \frac{v}{\Lambda}h_{1,2}
\end{equation}
where $v$ is the vev of the SM Higgs.

\subsection{Freeze-out regime}

\begin{figure}
    \centering
    \subfloat{\includegraphics[width=0.5\linewidth]{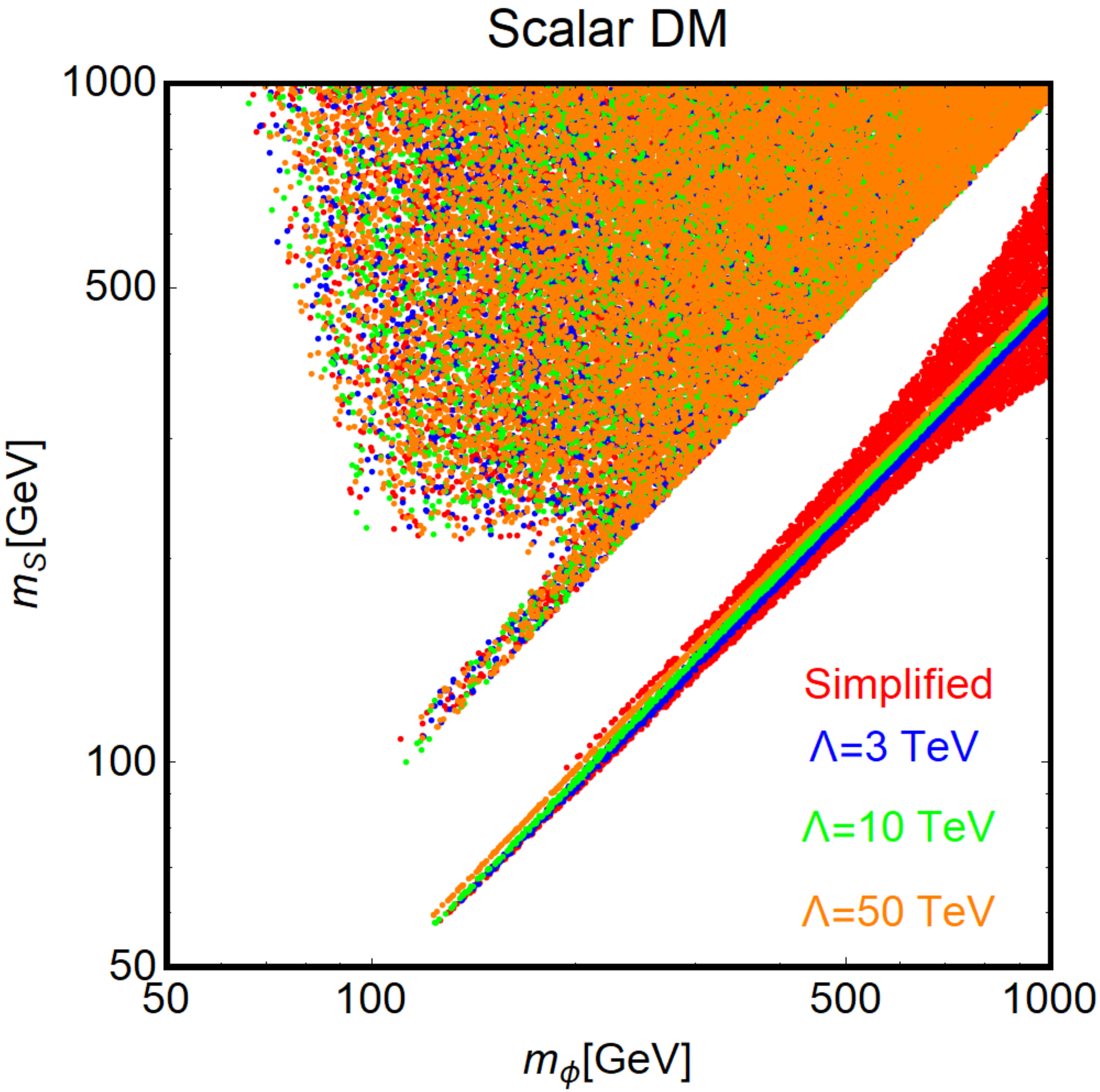}}
    \subfloat{\includegraphics[width=0.5\linewidth]{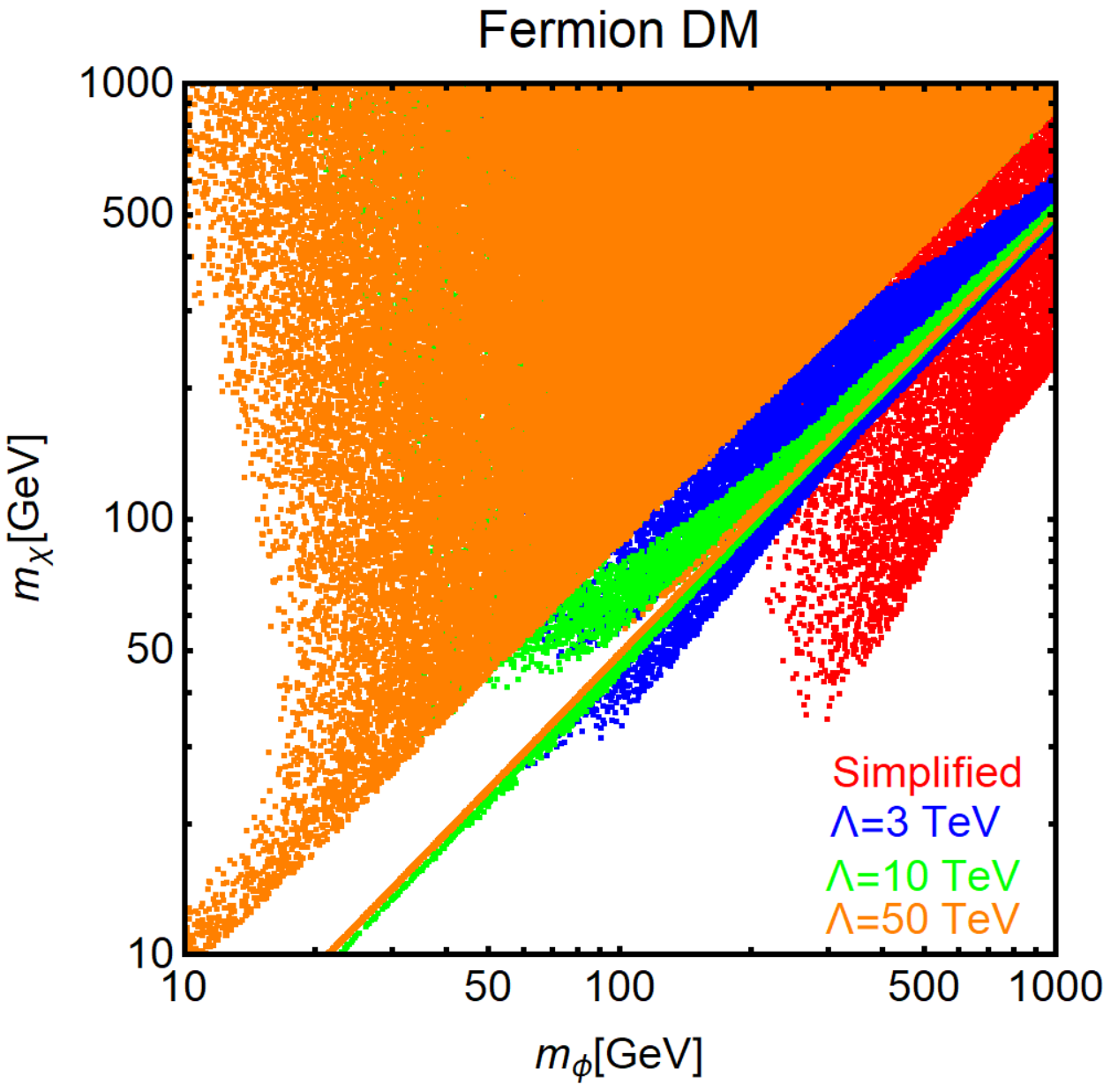}}
    \caption{Viable model points, in the $(m_{\rm DM},m_\phi)$ bidimensional plane, for scalar (left panel) and fermion (right panel) DM for different assignations of the NP scale $\Lambda$, reported in the panels themselves. For comparison, the plots includes also the viable model points (in red) associated to the simplified model depicted in section 3.}
    \label{fig:pmmL}
\end{figure}

Assuming again the thermal freeze-out as generation mechanism for the DM abundance, we can repeat the analysis performed in the previous section. There is, however, an additional parameter influencing DM phenomenology, namely the NP scale $\Lambda$ at which the flavor symmetry is broken. We have then repeated the parameter scan considering different assignations of the latter parameter, namely $\Lambda=3,10,50$ TeV. As shown in fig. \ref{fig:pmmL}, our findings in the $(m_{\chi,S},m_\phi)$-plane  have been compared with the results obtained in the previous section for the simplified model of eq.(\ref{hleadingorder}) (which would correspond to the case $\Lambda=246$ GeV).  

As evident, the viable parameter space, for $m_{S,\chi}<m_\phi$ progressively reduces as the scale $\Lambda$ increases. This is due to the $\frac{v}{\Lambda}$ suppression factor in the coupling between the mediator and the SM quarks. Because of this suppression, the DM annihilation cross-section into SM quarks cannot match the thermally favored values unless one relies on the resonant enhancement occurring for $m_{S,\chi}\sim \frac{m_\phi}{2}$. For $\Lambda > 50 \,\mbox{TeV}$, viable relic density is obtained only in the secluded regime, for $m_{S,\chi} > m_\phi$. In such a case, indeed, the relic density is determined essentially by the DM annihilation process into $\phi$ pairs, whose cross-section is independent from $\Lambda$. Consequently, DM evades most experimental searches ad exception of possible indirect signals from the $\text{DM} \,\text{DM} \rightarrow \phi \phi \rightarrow 4f$ process, mostly in the case of scalar DM.

In the secluded regime DM observables are not affected by specific value (and flavour structure) of the couplings of the mediator $\phi$ with the SM states. The only requirement is that these couplings are not too suppressed so that the DM could exist in thermal equilibrium in the Early Universe and then apply Standard Thermal freeze-out computations. As will be seen in the next subsection, this requirement can be used to set constraints on the scale $\Lambda$.

\subsection{Freeze-in regime}

As pointed in the previous subsection, the computation of the DM relic density, based on the thermal freeze-out, is based on the hypothesis that the DM was in thermal equilibrium, at least at temperatures higher than its mass. This might not be the case, however, if the interactions of the DM with the SM states are too suppressed. A rule of thumb to assess whether the DM particle was in thermal equilibrium in the Early Universe consists into comparing the DM annihilation rate $\Gamma_{\rm ann}=n_{\rm DM}^{\rm eq} \langle \sigma v \rangle$, with $n_{\rm DM}^{\rm eq}$ being the thermal equilibrium number density of the DM and $\langle \sigma v \rangle$ the thermally averaged annihilation cross-section into SM states, with the Hubble expansion rate, both computed at temperatures of the order of the DM mass. In the setup under consideration the size of the DM interactions with the SM primordial bath is mostly set by the scale $\Lambda$. We can then determine an upper bound on the scale $\Lambda$ of the form: 

\begin{align}
    & \left.\frac{\langle \sigma v \rangle n_\chi^{\rm eq}}{H}>1\right  \vert_{T=m_\chi} \nonumber\\
    & \rightarrow \left \{
    \begin{array}{cc}
    \Lambda \lesssim 6.3 \times 10^6\,\mbox{GeV} g_1 {\left(\frac{m_\chi}{100\,\mbox{GeV}}\right)}^{3/2}{\left(\frac{1\,\mbox{TeV}}{m_S}\right)}^2   & m_S \ll \frac{m_\phi}{2}  \\ 
    \Lambda \lesssim 2.2 \times 10^8 \,\mbox{GeV} g_1 {\left(\frac{100\,\mbox{GeV}}{m_\chi}\right)}^{1/2}   & m_S \gg \frac{m_\phi}{2}
    \end{array}
    \right.\nonumber\\
    & \rightarrow \left \{
    \begin{array}{cc}
    \Lambda \lesssim 3.5 \times 10^6\,\mbox{GeV} {\left(\lambda_2^2+\frac{3}{2}\lambda_1^2\right)}^{1/2} {\left(\frac{m_\chi}{100\,\mbox{GeV}}\right)}^{3/2}{\left(\frac{1\,\mbox{TeV}}{m_S}\right)}^2   & m_\chi \ll \frac{m_\phi}{2}  \\ \\
    \Lambda \lesssim 4.4 \times 10^7 \,\mbox{GeV} {\left(\lambda_2^2+\frac{3}{2}\lambda_1^2\right)}^{1/2} {\left(\frac{100\,\mbox{GeV}}{m_\chi}\right)}^{1/2}   & m_\chi \gg \frac{m_\phi}{2}
    \end{array}
    \right.
\end{align}
for scalar and fermionic DM, respectively. 
%\textbf{Cosa succede alla DM scalare?}.

Since the DM can also annihilate into mediator pairs, if the process is kinematically allowed, it can be maintained into thermal equilibrium as long as the mediator is. Having assumed $\mu_1=0$, the latter condition can be checked by comparing $H$ with the rate associated to the $\phi \leftrightarrow \bar q q$ process. From this we can infer the following condition on the scale $\Lambda$:
\begin{equation}
    \left. \frac{\langle \Gamma \rangle}{H} \right \vert_{T=m_\phi} \approx 1.29\, {\left(\frac{10^8 \,\mbox{GeV}}{\Lambda}\right)}\left(\frac{1\,\mbox{TeV}}{m_\phi}\right)\,.
\end{equation}

Even if the DM never achieved thermal equilibrium in the Early Universe, it can be efficiently produced, through the freeze-in mechanism \cite{Hall:2009bx}, from $\bar q q \rightarrow \bar \chi \chi (SS)$ annihilation processes. In such a case the DM relic density can be obtained by solving the following Boltzmann's equation, tracking the time evolution of the DM number density $n_{\rm DM}$:
\begin{equation}
    \frac{dn_{\rm DM}}{dt}+3Hn_{\rm DM}=\mathcal{N}(XX\rightarrow DM DM)\,,
\end{equation}
where $X$ is a SM state in thermal equilibrium (the sum over all the possible annihilation process is implicitly assumed). The right hand side of the equation is formally written as:
\begin{equation}
    \mathcal{N}(XX \rightarrow DM DM)=\frac{T g_X^2 |\eta_X|^2}{32\pi^4}\int ds s^{3/2}\sigma(s)\tilde{K}_1 (\sqrt{s}/T,x_X,x_X,0,\eta_X,\eta_X) \,,
\end{equation}
where $\sigma(s)$ is the annihilation cross-section as a function of the center of mass energy $s$ while the function $\tilde{K}_1$ accounts for the fact that in the freeze-in regime one has to adopt on Fermi-Dirac distribution functions, including a chemical potential, encoded in the parameter $\eta_i=\eta_s \exp\left(\mu_i/T\right)$ with $\eta_s=-1 (+1)$ for fermions (boson), for the SM fermions rather than relying on the Maxwell-Boltzmann distribution, as done in the case of WIMP production mechanism. Finally $x_X=m_X/T$ while $g_X$ represents the internal degrees of freedom of the SM state X. By doing the customary change of variables $n_{\rm DM} \rightarrow Y_{\rm DM}=n_{\rm DM}/s$, $\frac{d}{dt}\rightarrow -HT \frac{d}{dT}$ we can write the solution of the Bolzmann's equation as:
\begin{equation}
\label{eq:FIsol}
    Y_{\rm DM}=\frac{45}{64 \pi^4}\sqrt{\frac{90}{\pi^2}}M_{\rm Pl}g_X^2|\eta_X|^2 \int \frac{dT}{T^5}\frac{1}{\sqrt{g_{*\rho}}g_{*s}}\mathcal{N}(XX \rightarrow DM DM)\,,
\end{equation}
where the integral is computed between the present time temperature $T_0$ and the reheating temperature $T_R$. We remind that the reheating temperature is the temperature at which the radiation dominated epoch in standard cosmology begins. The latter has been assumed to be below the scale of breaking of the flavor symmetry, so that lagrangians \ref{eq:lagrangian_scalar}-\ref{eq:lagrangian_fermion} can be adopted for the computations. $g_{*,\rho}$ and $g_{*,s}$ represent, respectively, the number of effective relativistic degrees of freedom contributing to the energy density and the entropy density. Contrary to the case of freeze-out, we see that in the case of freeze-in production the DM relic density is proportional to the strength of the DM interactions with the SM states.

This integral, and the related DM relic density have been computed through the package micrOMEGAs\,5 \cite{Belanger:2018ccd}. We will nevertheless provide below some semi-analytical estimates in order to improve the understanding of the results.

\begin{figure}
    \centering
    \includegraphics[width=0.45\linewidth]{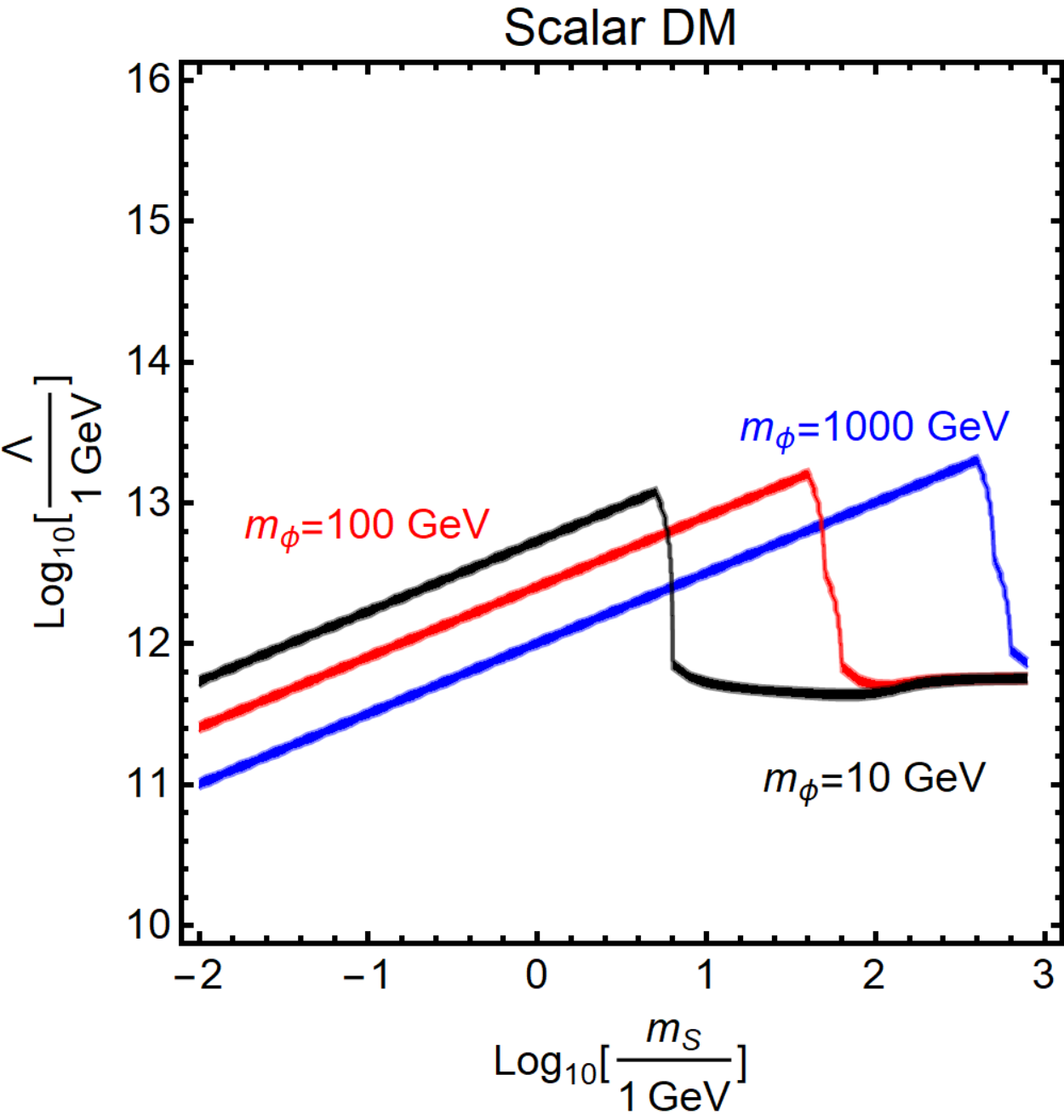}
    \includegraphics[width=0.45\linewidth]{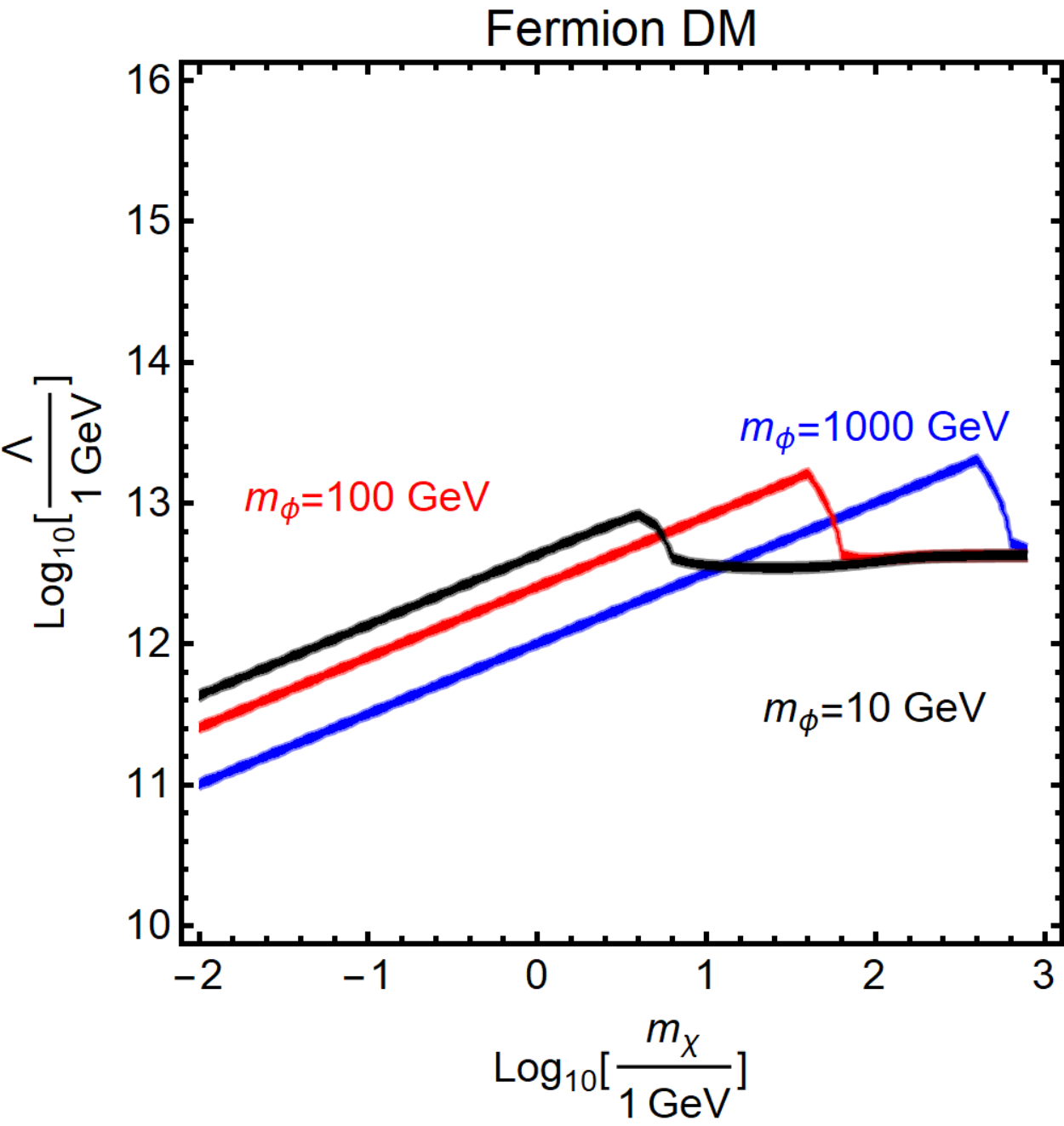}
    \caption{Isocontours of the correct DM relic density, assuming production through freeze-in, in the $(m_S,\Lambda)$ (left panel) and $(m_\chi,\Lambda)$ (right panel) bidimensional planes, for the three assignations $m_\phi=10,100,1000\,\mbox{GeV}$.}
    \label{fig:OMEGAFI}
\end{figure}

Fig \ref{fig:OMEGAFI} shows contours of the correct DM relic density in the $(m_S,\Lambda)$ (left panel) and $(m_\chi,\Lambda)$ (right panel) planes, for the three assignations $m_\phi=10,100,1000\,\mbox{GeV}$. In both scalar and fermionic DM cases, the relic density contours evidence two distinct trends. For $m_{\chi,S}\ll m_\phi$, the DM relic density increases with the DM mass, so that a comparable increase of the value of the NP scale $\Lambda$ is needed. In the opposite case, $m_{\chi,S} \gg m_\phi$, the DM relic density appears to be independent from both the DM and the mediator masses, being just set by the value of $\Lambda$. These two different regimes can be explained as follows.

Let us start with the case $m_{S,\chi} \gg m_\phi$. In such a case, for both fermionic and scalar DM, the cross section $\sigma(s)$ can be approximated as $\sigma(s)\simeq \frac{\kappa}{s}\frac{v^2}{\Lambda^2}$, where $\kappa$ is a parameter containing the couplings and numerical factors. We can at this point operate the following change of variables: $s\rightarrow z=\sqrt{s}/T$ and $T\rightarrow x=m_{\rm DM}/T$. In such a way eq. \ref{eq:FIsol} can be rewritten as:
\begin{equation}
    Y_{DM}=\frac{45}{32 \pi^4} \kappa\sqrt{\frac{90}{\pi^2}}\frac{M_{\rm Pl}}{m_{\rm DM}}\frac{v^2}{\Lambda^2}g_X^2 |\eta_X|^2\int dx \int dz z^2 \tilde{K}_1 (z,x_X,x_X,0,\eta_X,\eta_X)\,,
\end{equation}
where the argument of the integral does not depend explicitly on the masses of the new particles. Being $\Omega_{\rm DM}=m_{\rm DM}Y_{\rm DM}/(s_0 \rho_c)$, with $s_0$ being the entropy density at present times while $\rho_c$ is the so called critical density, it is easy to see that $\Omega_{\rm DM} \propto \frac{v^2}{\Lambda^2}$, hence without explicit dependence on the DM mass.

In the case in which, instead, the DM is sensitively lighter than the mediator, DM production is in the regime dubbed on-shell in \cite{Belanger:2018ccd}, in which the decay of the mediator into DM pairs is the most relevant effect for the DM relic density.
In such a case we have that:
\begin{equation}
    \sigma(s)=\frac{g_\phi}{g_X^2}\frac{4\pi^2 m_\phi}{(p_X^{\rm CM})^2}\frac{\Gamma(\phi \rightarrow XX) \Gamma(\phi \rightarrow DM DM)}{\Gamma_{\rm tot}}\delta(s-m_\phi^2)\,,
\end{equation}
which simply gives:
\begin{equation}
    \mathcal{N}(XX \rightarrow DM DM)=\frac{T g_\phi |\eta_X|^2}{2\pi^2}m_\phi^2  \frac{\Gamma(\phi \rightarrow XX) \Gamma(\phi \rightarrow DM DM)}{\Gamma_{\rm tot}}\tilde{K}_1(y,x_X,x_X,0,\eta_X,\eta_X)\,,
\end{equation}
where $y=m_\phi/T$. Using the latter as independent variable in place of the temperature we can write:
\begin{equation}
    Y_{DM}=\kappa \frac{45}{4 \pi^4}\sqrt{\frac{90}{\pi^2}}\frac{M_{\rm Pl}}{m_\phi}g_X^2|\eta_X|^2 \int dy y^3 \tilde{K}_1 (y,x_X,x_X,0,\eta_X,\eta_X)
\end{equation}
where we have posed $\Gamma(\phi \rightarrow XX)=\kappa m_\phi \frac{v^2}{\Lambda^2}$ and $\frac{\Gamma(\phi \rightarrow DM DM)}{\Gamma_{\rm tot}} \sim 1$ since the decay rate of the mediator into DM is not suppressed by the scale $\Lambda$. It is then immediate to see that the dependence of the DM relic density on the model parameters is of the form, $\Omega_{\rm DM} \propto \frac{m_{\rm DM}}{m_\phi}\frac{v^2}{\Lambda^2}$.

\section{Conclusion}
\label{sec:conclusion}

In this work we have illustrated the phenomenology of a dark model embedded in a $S_4 \times Z_5$ framework. Focusing at first on a simplified low energy model, in which the flavor symmetry has been just used as ansatz for the structure of the coupling with SM quark of a generic spin-0 mediator, we have shown that, in the cases of both scalar and fermionic DM, it is possible to achieve the correct DM relic density and at the same time comply with constraints from DM searches. Furthermore, we have seen that in the case of fermionic DM, there are specific viable regions of parameter space not present in other simplified models.
In a more concrete realization of the scenario under consideration, the DM interactions are suppressed by the scale $\Lambda$ associated to the breaking of the flavor symmetry. If the latter scale is above 10 (50) TeV for scalar (fermionic) DM, a viable phenomenology is obtained only in the so called secluded regime, in which the relic density is obtained mostly through annihilation in mediator pairs. We have finally considered the possibility in which the NP scale is very large, so that the DM was not existing in thermal equilibrium in the Early Universe. The correct relic density is nevertheless achieved through the freeze-in mechanism for a wide range of values of the DM and mediator masses and for $\Lambda \simeq 10^{11}-10^{12}\,\mbox{GeV}$.

%\section*{Acknowledgements}

\appendix

\section*{\bfseries Appendix}

\section{An $S_4 \times Z_5$ flavor symmetry realization}
\label{sec:S4_model}

To illustrate how the DM interaction in simplified models can be determined by 
a flavor symmetry we make use of a (slightly modified version of a) realistic model based on the $S_4 \times Z_5$ flavor symmetry, which has been shown to reproduce the flavor structure of the SM with a good accuracy~\cite{Meloni:2009cz}. 
The model assumes the existence of a number of new scalar fields whose transformation properties (together with the SM Higgs) are listed in \reftab{tab:transform}.
\begin{table*}[h!]
\centering
\begin{tabular}{cccccccc}
\toprule
 Fields& $h$& 
$\varphi_T$ & $\eta$ & $\Delta$ & $\varphi_S$ & $\xi$ \\
\headrule
$S_4$ & $1_1$ &$3_1$ & $2$ & $2$ & $3_1$ &  $1_1$ \\
% \hline
$Z_5$ & $1$&
$\omega^4$ & $\omega^4$  & $\omega$ & $\omega$ & $\omega$ \\
\bottomrule
\end{tabular}
\caption{\it Transformation properties of flavons and Higgs fields under the flavor symmetry $S_4 \times
Z_5$.}
\label{tab:transform}
\end{table*}
The flavor symmetry is broken at a generic large energy scale $\Lambda$ by the  vacuum expectation values (vevs) of such fields which, in flavor space, point along the following directions:
\begin{align}
\langle \eta \rangle &=v_\eta\,(0, 1)\,, &
\langle \varphi_T \rangle &= v_T\,(0,1,0) \,,
\label{solT} 
\end{align}
and
\begin{align}
\label{solS}
\langle \xi \rangle &=u\,, &
\langle \Delta \rangle &= v_\Delta\,(1,1)\,, &
\langle \varphi_S \rangle &=v_S\,(1, 1,1)\,.
\end{align} 
In the absence of any dynamical reason, we can assume the same order of magnitude for the vevs $v_i$ and introduce the order parameter $\varepsilon = \langle scalar
\rangle /\Lambda$ which, in the flavor model, governs the ratio of the charged lepton masses as well as the relevant NLO corrections to the neutrino mixing matrix that are needed to shift  the (simplistic) LO predictions to a matrix compatible with the current neutrino data \cite{Esteban:2018azc}. Thus, we expect $\varepsilon \sim 0.05$. For the quark sector, which is the relevant one for this paper,  we report in \reftab{tab:quarks1} the transformation properties under the flavor symmetry, where $Q$ is a triplet of $SU(2)$ quark doublets.

\label{quarks}
\begin{table*}[h!]
\centering
\begin{tabular}{ccccc}
\toprule
Field& $Q$ & $u^c,d^c$ & $c^c,s^c$ & $t^c,b^c$\\
\hline
$S_4$ & $3_1$ & $1_2$ & $1_1$ &$1_1$ \\
% \hline
$Z_5$ & 1 &  $\omega^3$& $\omega^2$ &  $\omega$\\
\bottomrule
\end{tabular}
\caption{\it Transformation properties of quarks under $S_4 \times Z_5$.}
\label{tab:quarks1}
\end{table*}
This assignment is enough to correctly reproduce the ratio among the down-type quarks while a small amount of fine-tuning in the Yukawa couplings is needed to accommodate the large top quark mass as well as the (13) and (23) entries of the CKM. This problem can be prevented by allowing a soft hierarchy among the vevs in eqs.(\ref{solT}) and (\ref{solS}) which, however, is a possibility not contemplated in the present paper.

In order to account for DM in our theory we introduce two states, a scalar (S) of fermionic ($\chi$) DM candidate and a mediator field $\phi$. All these new states are assumed to be singlet both under the SM gauge group and the flavor discrete symmetry. In a similar vein as \cite{Alanne:2020xcb} we can write the following lagrangian:
\begin{eqnarray}
\mathscr{L}&=&
\frac{y_b}{\Lambda^2} b^c (Q \varphi_T) \, h\,\phi +\nonumber\\
&&\frac{y_{s_1}}{\Lambda^3} s^c (Q\, \varphi_T \,\varphi_T) \, h\,\phi +
\frac{y_{s_2}}{\Lambda^3} s^c Q\, (\eta\,\varphi_T) \, h\,\phi + \nonumber\\
&&\nonumber\frac{y_{d_1}}{\Lambda^3} d^c \,Q\,[\varphi_T 
(\varphi_T\varphi_T)_2]_{3_2} \, h\,\phi +
\frac{y_{d_2}}{\Lambda^3} d^c \,Q\,[\varphi_T (\varphi_T\varphi_T)_{3_1}]_{3_2} 
\, h\,\phi +\\
&&\nonumber
\frac{y_{d_3}}{\Lambda^4} d^c \,Q\, [\eta\, (\varphi_T\varphi_T)_{3_1}] 
_{3_2}\, 
h\,\phi 
+ \frac{y_{d_4}}{\Lambda^4} d^c \,Q\, [\varphi_T \, (\eta\eta)_{2}] _{3_2}\, 
h\,\phi +\\
&&\label{lagquark}
\frac{y_{d_5}}{\Lambda^3} d^c \,Q\,  (\Delta \varphi_S)_{3_2}\, h\,\phi\,.
\end{eqnarray}
For simplicity we will not consider the possibility of mass mixing between the $\phi$ and $h$ states. In addition we will assume isoscalar interactions, \ie, equality between the 
couplings of up and down quarks in each generation. 

After flavor and electroweak symmetry breaking the previous Lagrangian 
generates an effective term that corresponds to the term $q^c \, h_1 \,q$
of \refeq{eq:Sphi} and \refeq{eq:phichi}, with
\begin{equation}
h_1=\frac{v}{\Lambda}\left(
\begin{array}{ccc} a_1 \, \varepsilon^3 & 0 & 0 \\
0 & b_2\,\varepsilon^2 & 0\\
0 & 0 & c_3\,\varepsilon
\end{array}
\right)\,,
\label{mquarkfin}
\end{equation}
where the coefficients $a_i,b_i,c_i$ are linear combinations of the $y_i$ 
parameters and the Higgs VEVs, and $\varepsilon = \langle \varphi 
\rangle /\Lambda$, assuming a common order of magnitude of the flavon vevs. 
%Unless differently stated we will assume the fixed value $\epsilon=0.05$, in the correct ballpark to reproduce the structure of the SM fermion mass matrices.
While the $h_1$ matrix is flavor diagonal at the leading order, all its entries become not null once the next to leading order effects from the corrections to the vacuum alignment of the flavon fields and from  
higher order operators of the form:
\begin{eqnarray}
\label{newc}
(\varphi_S^3),\,(\varphi_S\Delta^2),\,(\varphi_S\xi^2),\,(\varphi_S\Delta\xi)\,.
\end{eqnarray}
are taken into account:
%The resulting coupling matrix is as follows:
\begin{equation}
h_1=\frac{v}{\Lambda}\left(
\begin{array}{ccc} a_1 \, \varepsilon^3 &a_2 \,\varepsilon^3& -a_2 
\,\varepsilon^3\\
b_1\,\varepsilon^3 & b_2\,\varepsilon^2 & b_3\,\varepsilon^3\\
c_1\,\varepsilon^2 & c_2\,\varepsilon^2 & c_3\,\varepsilon
\end{array}
\right)\,,
\label{mquarkfin2}
\end{equation}
Being of higher order in the $\epsilon \ll 1$ parameter, the off-diagonal entries of the $h_1$ matrix have a negligible impact in DM phenomenology. For the latter it is then enough to just take its leading order expression. The coefficients $a_{1},b_{2},c_{3}$ are combinations of SM Yukawa couplings and are expected to be of order 1. For simplicity we will assign $a_1 = b_2 = c_3 = 1$ throughout our study.

\section{The Group $S_4$}
\label{sec:app_S4}
The structure of $h_{1,2}$ used in our numerical simulations has been obtained adopting the following convention for the generators $S$ and $T$, according to: 
\begin{equation}
S^4= T^3=  (ST^2)^2=\unity\,.
\end{equation}
In the different representations, they can be written as reported in 
\reftab{tab:app_repre}.
\begin{table}[t!]
\centering
\begin{tabular}{cccccc}
\toprule
Representation & $1_1$ &$1_2$ & $2$ & $3_1$ & $3_2$ \\   \headrule   
$S$  & 1  &  -1 & $\left(\begin{array}{cc} 0 & 1 \\ 1 & 0 \\ \end{array} 
\right)$  & $\dfrac{1}{3}\left(
                                 \begin{array}{ccc}
                                   -1 & 2\omega & 2\omega^2 \\
                                   2\omega & 2\omega^2 & -1 \\
                                   2\omega^2 & -1 & 2\omega \\
                                 \end{array}
                               \right)$  & $\dfrac{1}{3}\left(
                                 \begin{array}{ccc}
                                   1 & -2\omega & -2\omega^2 \\
                                   -2\omega & -2\omega^2 & 1 \\
                                   -2\omega^2 & 1 & -2\omega \\
                                 \end{array}
                               \right)$ \\

$T$ & 1  & 1 & $\left(\begin{array}{cc} \omega & 0 \\ 0 & \omega^2 \\ 
\end{array} 
\right)$  &  $\left(
                                              \begin{array}{ccc}
                                                1 & 0 & 0 \\
                                                0 & \omega^2 & 0 \\
                                                0 & 0 & \omega \\
                                              \end{array}
                                            \right)$ &$\left(
                                              \begin{array}{ccc}
                                                1 & 0 & 0 \\
                                                0 & \omega^2 & 0 \\
                                                0 & 0 & \omega \\
                                              \end{array}
                                            \right)$ \\
\bottomrule
\end{tabular}
\caption{Generators $S$ and $T$ in different representations. 
\label{tab:app_repre}}
\end{table}
In the previous basis, the Clebsch-Gordan coefficients are as follows
($\alpha_i$ indicates the elements of the first
representation of the product and $\beta_i$ the second one):
\[
\begin{array}{lcl}
1_1\otimes\eta&=&\eta\otimes1_1=\eta\quad\text{with $\eta$ any
representation}\\[-10pt]
\\[8pt]
1_2\otimes1_2&=&1_1\sim\alpha\beta\\[-10pt]
\\[8pt]
1_2\otimes2&=&2\sim\left(\begin{array}{c}
                    \alpha\beta_1 \\
                    -\alpha\beta_2 \\
            \end{array}\right)\\[-10pt]
\\[8pt]
1_2\otimes3_1&=&3_2\sim\left(\begin{array}{c}
                    \alpha\beta_1 \\
                    \alpha\beta_2 \\
                    \alpha\beta_3 \\
                    \end{array}\right)\\[-10pt]
\\[8pt]
1_2\otimes3_2&=&3_1\sim\left(\begin{array}{c}
                            \alpha\beta_1 \\
                            \alpha\beta_2 \\
                            \alpha\beta_3 \\
                    \end{array}\right) \,.
\end{array}
\]
The multiplication rules with the 2-dimensional
representation are the following:
\[
\begin{array}{ll}
2\otimes2=1_1\oplus1_2\oplus2&\quad
\text{with}\quad\left\{\begin{array}{l}
                    1_1\sim\alpha_1\beta_2+\alpha_2\beta_1\\[-10pt]
                    \\[8pt]
                    1_2\sim\alpha_1\beta_2-\alpha_2\beta_1\\[-10pt]
                    \\[8pt]
                    2\sim\left(\begin{array}{c}
                        \alpha_2\beta_2 \\
                        \alpha_1\beta_1 \\
                    \end{array}\right)
                    \end{array}
            \right.\\[-10pt]
\\[8pt]
2\otimes3_1=3_1\oplus3_2&\quad
\text{with}\quad\left\{\begin{array}{l}
                    3_1\sim\left(\begin{array}{c}
                        \alpha_1\beta_2+\alpha_2\beta_3 \\
                        \alpha_1\beta_3+\alpha_2\beta_1 \\
                        \alpha_1\beta_1+\alpha_2\beta_2 \\
                    \end{array}\right)\\[-10pt]
                    \\[8pt]
                    3_2\sim\left(\begin{array}{c}
                        \alpha_1\beta_2-\alpha_2\beta_3\\
                        \alpha_1\beta_3-\alpha_2\beta_1 \\
                        \alpha_1\beta_1-\alpha_2\beta_2 \\
                    \end{array}\right)\\
                    \end{array}
            \right.\\[-10pt]
\\[8pt]
2\otimes3_2=3_1\oplus3_2&\quad
\text{with}\quad\left\{\begin{array}{l}
                    3_1\sim\left(\begin{array}{c}
                        \alpha_1\beta_2-\alpha_2\beta_3\\
                        \alpha_1\beta_3-\alpha_2\beta_1 \\
                        \alpha_1\beta_1-\alpha_2\beta_2 \\
                    \end{array}\right)\\[-10pt]
                    \\[8pt]
                    3_2\sim\left(\begin{array}{c}
                        \alpha_1\beta_2+\alpha_2\beta_3 \\
                        \alpha_1\beta_3+\alpha_2\beta_1 \\
                        \alpha_1\beta_1+\alpha_2\beta_2 \\
                    \end{array}\right)\\
                    \end{array}
            \right.\\
\end{array}
\]
The multiplication rules with the 3-dimensional
representations are the following:
\[
\begin{array}{ll}
3_1\otimes3_1=3_2\otimes3_2=1_1\oplus2\oplus3_1\oplus3_2\qquad
\text{with}\quad\left\{
\begin{array}{l}
1_1\sim\alpha_1\beta_1+\alpha_2\beta_3+\alpha_3\beta_2 \\[-10pt]
                    \\[8pt]
2\sim\left(
     \begin{array}{c}
       \al_2\beta_2+\al_1\beta_3+\al_3\beta_1 \\
       \al_3\beta_3+\al_1\beta_2+\al_2\beta_1 \\
     \end{array}
   \right)\\[-10pt]
   \\[8pt]
3_1\sim\left(\begin{array}{c}
         2\al_1\beta_1-\alpha_2\beta_3-\alpha_3\beta_2 \\
         2\al_3\beta_3-\alpha_1\beta_2-\alpha_2\beta_1 \\
         2\al_2\beta_2-\alpha_1\beta_3-\alpha_3\beta_1 \\
        \end{array}\right)\\[-10pt]
        \\[8pt]
3_2\sim\left(\begin{array}{c}
         \alpha_2\beta_3-\alpha_3\beta_2 \\
         \alpha_1\beta_2-\alpha_2\beta_1 \\
         \alpha_3\beta_1-\alpha_1\beta_3 \\
    \end{array}\right)
\end{array}\right.
\end{array}
\]
\[
\begin{array}{ll}
3_1\otimes3_2=1_2\oplus2\oplus3_1\oplus3_2\qquad
\text{with}\quad\left\{
\begin{array}{l}
1_2\sim\alpha_1\beta_1+\alpha_2\beta_3+\alpha_3\beta_2\\[-10pt]
        \\[8pt]
2\sim\left(
     \begin{array}{c}
       \al_2\beta_2+\al_1\beta_3+\al_3\beta_1 \\
       -\al_3\beta_3-\al_1\beta_2-\al_2\beta_1 \\
     \end{array}
   \right)\\[-10pt]
        \\[8pt]
3_1\sim\left(\begin{array}{c}
         \alpha_2\beta_3-\alpha_3\beta_2 \\
         \alpha_1\beta_2-\alpha_2\beta_1 \\
         \alpha_3\beta_1-\alpha_1\beta_3 \\
    \end{array}\right)\\[-10pt]
        \\[8pt]
3_2\sim\left(\begin{array}{c}
         2\al_1\beta_1-\alpha_2\beta_3-\alpha_3\beta_2 \\
         2\al_3\beta_3-\alpha_1\beta_2-\alpha_2\beta_1 \\
         2\al_2\beta_2-\alpha_1\beta_3-\alpha_3\beta_1 \\
    \end{array}\right)\\
\end{array}\right.
\end{array}
\]

\bibliographystyle{modified}
\bibliography{refs}

% \end{multicols}

\end{document}